\documentclass[twocolumn]{aastex7}
\usepackage{url}
\usepackage{xspace}
\let\oldAA\AA
\renewcommand{\AA}{{\text{\normalfont\oldAA}}}
\newcommand{\eg}{\textit{e.g.}}

\usepackage{xcolor}
\newcommand{\target}{\object{Cosmic Owl}\xspace}
\newcommand{\CI}{[CI]($\rm^3P_1$-$\rm^3P_0$)\xspace}

\shorttitle{The Cosmic Owl}
\shortauthors{M. Li et al.}
\graphicspath{{./}{figures/}}

\begin{document}

\title{
The Cosmic Owl: Twin Active Collisional Ring Galaxies with Starburst Merging Front at $z=1.14$
}

\author[0000-0001-6251-649X]{Mingyu Li}
\affiliation{Department of Astronomy, Tsinghua University, Beijing 100084, China}
\email[show]{lmytime@hotmail.com}

\author[0000-0003-2983-815X]{Bjorn\,H.\,C. Emonts}
\affiliation{National Radio Astronomy Observatory, 520 Edgemont Road, Charlottesville, VA 22903, USA}
\email{bemonts@nrao.edu}

\author[0000-0001-8467-6478]{Zheng Cai}
\affiliation{Department of Astronomy, Tsinghua University, Beijing 100084, China}
\email[show]{zcai@mail.tsinghua.edu.cn}

\author[orcid=0009-0003-4742-7060]{Takumi S. Tanaka}
\affiliation{Kavli Institute for the Physics and Mathematics of the Universe (WPI), The University of Tokyo Institutes for Advanced Study, The University of Tokyo, Kashiwa, Chiba 277-8583, Japan}
\affiliation{Department of Astronomy, Graduate School of Science, The University of Tokyo, 7-3-1 Hongo, Bunkyo-ku, Tokyo 113-0033, Japan}
\affiliation{Center for Data-Driven Discovery, Kavli IPMU (WPI), UTIAS, The University of Tokyo, Kashiwa, Chiba 277-8583, Japan}
\email{takumi.tanaka@ipmu.jp}

\author[0000-0001-6865-499X]{Wilfried Mercier}
\affiliation{Aix Marseille Univ., CNRS, CNES, LAM, Marseille, France}
\email{wilfried.mercier@lam.fr}

\author[0000-0003-0111-8249]{Yunjing Wu}
\affil{Department of Astronomy, Tsinghua University, Beijing 100084, China}
\email{yj-wu19@mails.tsinghua.edu.cn}

\author[0000-0002-3489-6381]{Fujiang Yu}
\affil{Department of Astronomy, Tsinghua University, Beijing 100084, China}
\email{yufj@mail.tsinghua.edu.cn}

\author[0000-0002-4622-6617]{Fengwu Sun}
\affiliation{Center for Astrophysics $|$ Harvard \& Smithsonian, 60 Garden St., Cambridge, MA 02138, USA}
\email{fengwu.sun@cfa.harvard.edu}

\author[0000-0002-1620-0897]{Fuyan Bian}
\affiliation{European Southern Observatory, Alonso de C\'{o}rdova 3107, Casilla 19001, Vitacura, Santiago 19, Chile}
\email{Fuyan.Bian@eso.org}

\author[0000-0002-3331-9590]{Emanuele Daddi}
\affiliation{AIM, CEA, CNRS, Universit\'{e} Paris-Saclay, Universit\'{e} Paris Diderot, Sorbonne Paris Cit\'{e}, F-91191 Gif-sur-Yvette, France}
\email{emanuele.daddi@cea.fr}

\author[0000-0003-3310-0131]{Xiaohui Fan}
\affiliation{Steward Observatory, University of Arizona, 933 N Cherry Avenue, Tucson, AZ 85721, USA}
\email{xiaohuidominicfan@gmail.com}

\author[0000-0001-6052-4234]{Xiaojing Lin}
\affiliation{Department of Astronomy, Tsinghua University, Beijing 100084, China}
\affil{Steward Observatory, University of Arizona, 933 N Cherry Avenue, Tucson, AZ 85721, USA}
\email{linxj21@mails.tsinghua.edu.cn}

\author[0000-0002-6221-1829]{Jianwei Lyu}
\affiliation{Steward Observatory, University of Arizona, 933 N Cherry Avenue, Tucson, AZ 85721, USA}
\email{jianwei@arizona.edu}

\author[0000-0001-9187-3605]{Jeyhan S. Kartaltepe}
\affiliation{Laboratory for Multiwavelength Astrophysics, School of Physics and Astronomy, Rochester Institute of Technology, 84 Lomb Memorial Drive, Rochester, NY 14623, USA}
\email{jeyhan@astro.rit.edu}

\author[0000-0001-6477-4011]{Francesco Valentino}
\affiliation{Cosmic Dawn Center (DAWN), Denmark}
\affiliation{DTU Space, Technical University of Denmark, Elektrovej 327, DK2800 Kgs. Lyngby, Denmark}
\email{fmava@dtu.dk}


\begin{abstract}
Galaxy mergers play a critical role in driving galaxy evolution, especially by transforming galaxy morphology, redistributing gas around galaxies, triggering active galactic nuclei (AGN), and stimulating star formation. We present the Cosmic Owl, a galaxy merger at $z=1.14$, identified in the COSMOS field. Deep imaging and spectroscopy from JWST, ALMA, and VLA reveal a complex system of twin collisional ring galaxies, exhibiting nearly identical morphologies. The grism spectra from the JWST COSMOS-3D program confirm that both galaxies host an AGN. A bipolar radio jet from one AGN extends to strike the merging front. In addition, we detect a starburst at the merging front, characterized by luminous extended nebular line emission and a massive cold gas reservoir. This starburst is likely triggered by interstellar shocks induced by galaxy collision and the AGN jet. The twin ring structure of the Cosmic Owl requires further numerical simulations to clarify the precise conditions that lead to the formation of this rare morphology.  This system exemplifies how shock-induced star formation, driven by galaxy collision or AGN jet, can act as a crucial mechanism for triggering intense starbursts in the early Universe.
\end{abstract}



\section{Introduction}
\label{sec:intro}
Galaxy mergers are believed to drive the evolution of galaxies \citep{White1978MNRAS.183..341W,Toomre1972ApJ...178..623T}, serving as pivotal events in the hierarchical paradigm of structure formation across cosmic history \citep{White1978MNRAS.183..341W}.
Mergers can redistribute the gas around galaxies \citep[\eg,][]{Hani2018MNRAS.475.1160H}, impact the stellar kinematics \citep[\eg,][]{Clauwens2018MNRAS.478.3994C}, transform galaxy morphology \citep[\eg,][]{Dubois2016MNRAS.463.3948D, Rodriguez-Gomez2017MNRAS.467.3083R}, and eventually lead to effective stellar mass assembly \citep[\eg,][]{Rodriguez-Gomez2016MNRAS.458.2371R,Mundy2017MNRAS.470.3507M,Duncan2019ApJ...876..110D,Martin2021MNRAS.500.4937M}.

In certain configurations, mergers can lead to the formation of collisional ring galaxies \citep[CRGs; see][for a review]{Appleton1996FCPh...16..111A}.
These rings arise when one galaxy passes directly through the disk of another in a nearly head-on collision, causing gas and stars to be shocked outward into a circular or near-circular pattern.
Ring galaxies are exceptionally rare; only a few hundred are known in the local Universe \citep{Madore2009ApJS..181..572M,Pasha2025ApJ...980L...3P}, emphasizing the uniqueness of these events due to the precise alignment and interaction conditions needed for their formation.

Mergers are expected to trigger a significant starburst in gas-rich environments \citep[\eg,][]{Cortijo-Ferrero2017A&A...607A..70C,Patton2020MNRAS.494.4969P}.
Concurrently, the merger can funnel gas towards the central regions, enhancing accretion onto the central supermassive black holes (SMBHs), thereby igniting or intensifying active galactic nuclei (AGN) activity \citep[\eg,][]{Springel2005MNRAS.361..776S,Hopkins2006ApJS..163....1H,DiMatteo2012ApJ...745L..29D,Satyapal2014MNRAS.441.1297S,Ellison2019MNRAS.487.2491E}.
Physical processes triggered by the merger can further impact galaxy formation.
For instance, relativistic jets emanating from the AGN can interact with the interstellar medium (ISM), generate shock fronts that propagate through the galaxy, and result in complex feedback mechanisms to significantly influence galaxy evolution \citep{Hardcastle2020NewAR..8801539H,Cielo2018MNRAS.477.1336C,Morganti2017FrASS...4...42M,Fabian2012ARA&A..50..455F}.
On the one hand, shocks heat the surrounding gas, potentially suppressing star formation by expelling or warming the cold gas \citep[\eg,][]{Talbot2024MNRAS.528.5432T}.
On the other hand, shocks compress gas clouds, promoting gas cooling and condensation, thus leading to star formation \citep[\eg,][]{Croft2006ApJ...647.1040C}.
This dual role of AGN jets — both quenching and stimulating star formation — underscores the intricate feedback processes involved.

Understanding the individual and collective impacts of these physical processes occurring in mergers requires comprehensive multi-wavelength observations to disentangle their complex interplay.
However, at high-$z$, understanding the role of mergers becomes increasingly challenging due to the limited sensitivity and resolution of observational facilities.
The James Webb Space Telescope (JWST) can address this gap, with its outstanding sensitivity and spatial resolution to reveal the detailed structure of galaxies at high redshift, shedding light on both gaseous and stellar components.

\begin{figure*}
    \centering
    \includegraphics[width=\linewidth]{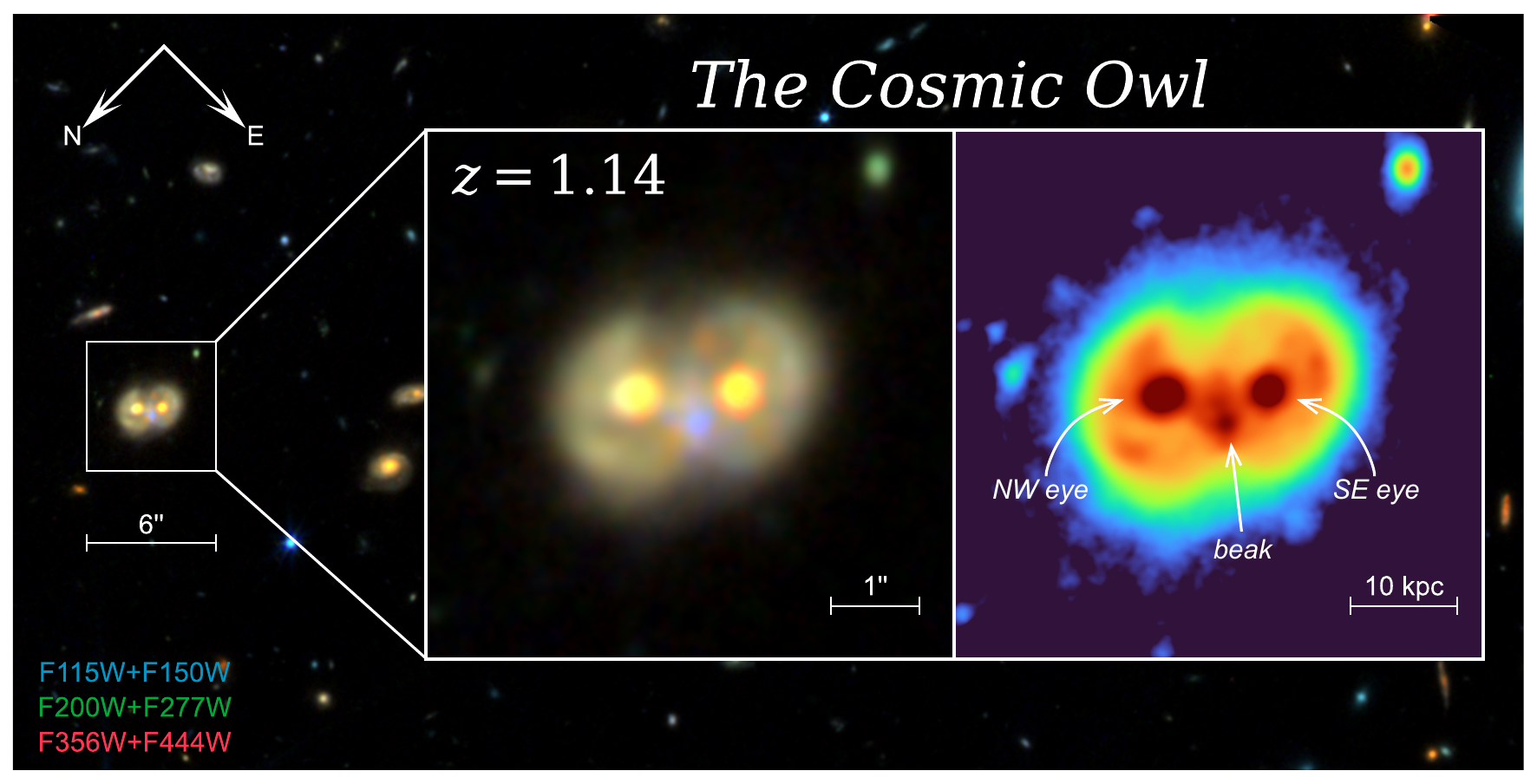}
    \includegraphics[width=\linewidth]{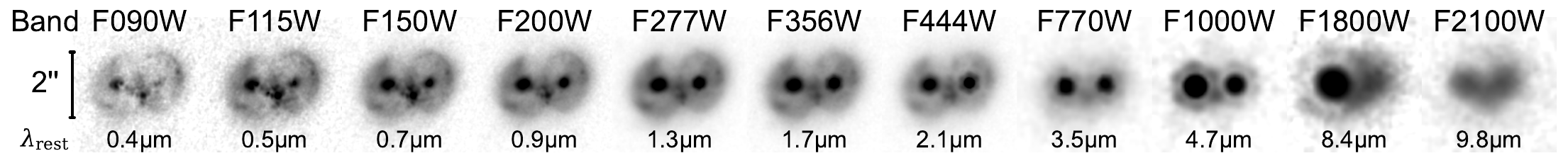}
    \caption{JWST NIRCam and MIRI wide band imaging of the Cosmic Owl.
    \textbf{Upper}: Pseudo-color image of the Cosmic Owl, with blue (F115W, F150W), green (F200W, F277W), and red (F356W, F444W) bands used. The right panel presents the composite image of all NIRCam wide bands (from F090W to F444W) after PSF matching. The three components, including \textit{NW eye}, \textit{SE eye}, and the \textit{beak}, are annotated in the right panel.
    \textbf{Lower}: Imaging cutout stamps for the Cosmic Owl, including seven NIRCam bands (F090W, F115W, F150W, F200W, F277W, F356W, and F444W) and four MIRI bands (F770W, F1000W, F1800W, and F2100W). The rest-frame wavelength relative to $z=1.14$ is annotated at the bottom of each panel. The twelve NIRCam and MIRI images, including the F410M data now shown, are available as the data behind the Figure in the online Journal.}
    \label{fig:cover}
\end{figure*}

In this paper, we present the serendipitous discovery of a unique galaxy system at $z=1.14$.
Due to its morphological similarities to an owl face (see Fig.~\ref{fig:cover}), we will hereafter refer to it as the \target.
The \target consists of a head-on merger involving two galaxies, each hosting an obscured AGN and exhibiting a collisional ring morphology.
In $\S$\ref{sec:obs}, we describe observations and data reduction procedures.
In $\S$\ref{sec:result}, we present the data analysis and results and we explore possible implications of these results in $\S$\ref{sec:discussion}, .
Finally, we summarize our major results in $\S$\ref{sec:summary}.
Throughout this work, magnitudes are given in the AB system \citep{Oke1983ApJ...266..713O}.
We adopt a flat $\mathrm{\Lambda CDM}$ cosmology with $\Omega_{\mathrm{m}}=0.3$ and $H_{0}=70 \mathrm{~km}\mathrm{~s}^{-1} \mathrm{~Mpc}^{-1}$.
In this cosmology, 1\arcsec\ corresponds to 8.225 kpc of physical length at $z=1.14$.

\section{Observations} \label{sec:obs}
The \target is situated within the COSMOS-CANDELS field \citep{Koekemoer2011ApJS..197...36K,Grogin2011ApJS..197...35G}, which constitutes the central region of the COSMOS extragalactic deep field \citep{Scoville2007ApJS..172....1S}, with extensive multiwavelength observations and well-studied galaxy catalogs.
The equatorial coordinates of this target are $\rm R.A.=150.0591202~deg$, $\rm Decl.=2.2199215~deg$ or $\rm \alpha=10^{h}00^{m}14^{s}.189, \delta=+02^{d}13^{m}11^{s}.72$.
We combine the following datasets to characterize this exceptional system: JWST, the Atacama Large Millimeter/submillimeter Array (\textit{ALMA}), the Karl G. Jansky Very Large Array (\textit{VLA}), and other multiwavelength observations from X-ray to radio.

\textbf{JWST imaging and spectroscopy}. The \target is observed by three JWST programs: COSMOS-Web \citep[GO\#1727; PIs: J. Kartaltepe and C. Casey;][]{Casey2023ApJ...954...31C}, PRIMER (Public Release IMaging for Extragalactic Research, GO\#1837; PI: J. Dunlop), and COSMOS-3D (GO\#5893; PI: K. Kakiichi), using the Near-Infrared Camera \citep[NIRCam;][]{Rieke2023PASP..135b8001R} with filter bands of F090W, F115W, F150W, F200W, F277W, F356W, F410M, F444W and the Mid-Infrared Instrument \citep[MIRI;][]{Wright2023PASP..135d8003W} with F770W, F1000W, F1800W, and F2100W bands.
The exposure time ranges from 1675s to 2786s per band.
To maintain consistency of the pipeline reference file,  ensure a consistent astrometric solution, and coadd the imaging in the same band from the different programs, we reduce all JWST data using the standard JWST pipeline\footnote{\url{https://github.com/spacetelescope/jwst}} version v1.18 \citep{bushouse_2025_15178003} with the recommended reference file of \texttt{jwst\_1364.pmap}.
We applied several customized data reduction steps, including 1/f noise removal, wisp removal, sky background subtraction, and astrometric correction.
The 1/f noise was subtracted in both the column and row directions, and wisps were removed using the public v3 templates provided by STScI.
The sky background subtraction was a two-stage process: first, for MIRI imaging, we subtracted a ``super sky'' created by stacking all single exposures within a mosaic visit; then, a global median background, derived after masking bright sources, was subtracted from each individual NIRCam and MIRI exposure.
The astrometry of each image is corrected to an absolute astrometry error of much smaller than 0.03 arcsec using the COSMOS2020 catalog \citep{Weaver2022ApJS..258...11W}, which has been registered to \textit{Gaia} data \citep{GaiaCollaboration2016A&A...595A...2G}.
The final mosaic images are drizzled with \texttt{pix\_frac=1.0} and a pixel size of 0\farcs03 for NIRCam and 0\farcs1 for MIRI.
The Cosmic Owl is observed by COSMOS-3D with NIRCam wide field slitless spectroscopy (WFSS) in the F444W band.
The F444W grism spectra cover the 3.9 -- 4.6~$\mu$m with a spectral resolution of $R\sim1600$.
The WFSS data are reduced following the routine shown by \citet{Sun2023ApJ...953...53S} and the pipeline codes are publicly available\footnote{\url{https://github.com/fengwusun/nircam_grism}} \citep{sun_2024_14052875}.

\textbf{\textit{ALMA} observations}. The \target is observed by \textit{ALMA} Band-6 in Cycle-4 (Project ID: 2016.1.01040.S; PI: F. Valentino) for CO(4-3) and \CI emission lines \citep{Valentino2018ApJ...869...27V}.
The \textit{ALMA} observations are taken in the configuration C40-1, corresponding to a synthesized beam of $\sim2\arcsec$.
The on-source integration time is 10.0 minutes.
The raw data are calibrated with the Common Astronomy Software Applications (CASA) v4.7.2 \citep{CASA2022PASP..134k4501C}, which includes the ALMA Cycle-4 pipeline \citep{Hunter2023PASP..135g4501H}.
The command \texttt{tclean} of CASA was used to perform the imaging with natural weighting, hogbom deconvolver, 5$\sigma$ rms threshold, and standard gridder.
The final cube has a channel width of 60~km/s in radio velocity.
The median root-mean-square (RMS) noise per 60~km~s$^{-1}$ channel is 0.62\,mJy\,beam$^{-1}$.

\textbf{\textit{VLA} radio imaging}. The \target is observed at radio wavelengths of 10 cm (3 GHz), 20 cm (1.4 GHz), and 90 cm (324 MHz) with the \textit{VLA} by the VLA-COSMOS 3 GHz Large Project \citep[\eg,][]{Smolcic2017A&A...602A...1S}, the 1.4 GHz Large Project \citep[\eg,][]{Schinnerer2007ApJS..172...46S}, and the 90 cm survey \citep[\eg,][]{Smolcic2014MNRAS.443.2590S}.
We use the publicly available mosaic-combined images from these surveys\footnote{https://irsa.ipac.caltech.edu/data/COSMOS/images/vla}.
These radio imaging observations have typical rms of 2~$\mu$Jy/beam, 10.5~$\mu$Jy/beam, 0.5~mJy/beam, and a typical beam (diameter) size of 0\farcs75\arcsec, 1\farcs5\arcsec, $8\arcsec\times6\arcsec$ at 10 cm, 20 cm, and 90 cm, respectively.

\textbf{Other archival data}. In addition to the data described above, we take advantage of the rich multi-wavelength imaging data available across the COSMOS field.
We utilize the multi-slit spectroscopic data obtained by Keck/DEIMOS, which are presented in the Appendix.
We use the photometry measurement from the COSMOS2020 catalog \citep{Weaver2022ApJS..258...11W}, which includes a wealth of ground-based optical and near-infrared (OIR) imaging data, including Hyper Suprime-Cam imaging \citep[HSC-SSP,][]{Aihara2022PASJ...74..247A}, CFHT u-band imaging \citep[CLAUDS survey,][]{Sawicki2019MNRAS.489.5202S}, and GALEX near-UV imaging \citep{Zamojski2007ApJS..172..468Z}.
Moreover, we also utilize X-ray data from the Chandra COSMOS Legacy Survey \citep{Civano2016ApJ...819...62C} and the infrared photometry measurements from ``Super-deblended'' catalogs \citep{Jin2018ApJ...864...56J}, including Spitzer/MIPS 24~$\mu$m \citep{Sanders2007ApJS..172...86S}.

\section{Analyses and Results}\label{sec:result}

\subsection{Twin Collisional Ring Galaxies}\label{ssec:ring}

\begin{figure*}[t]
    \centering
    \includegraphics[width=\linewidth]{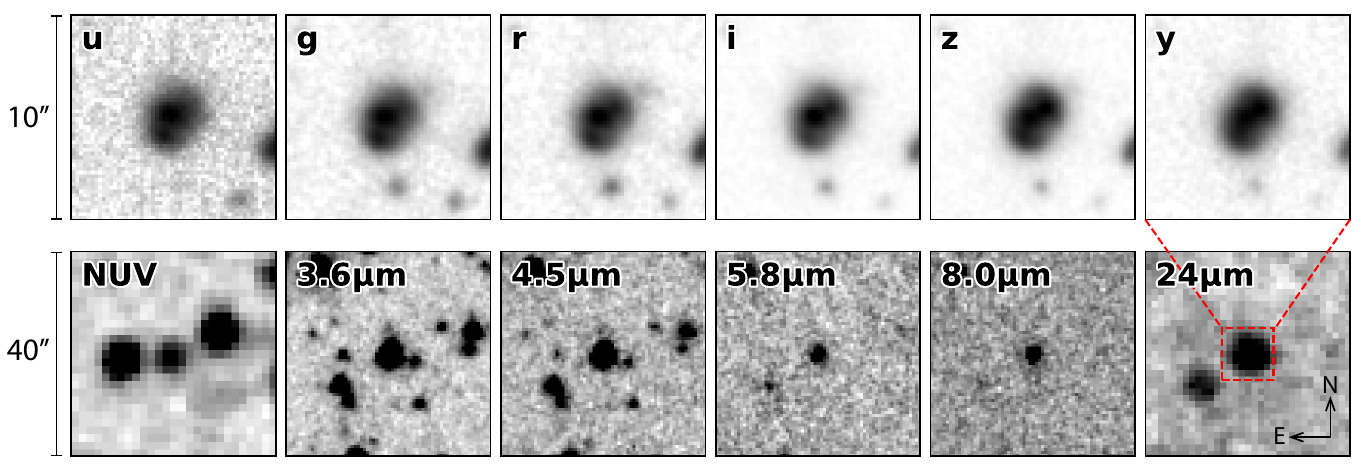}
    \caption{Postage stamp images utilized for the SED analysis of the Cosmic Owl, complementing the JWST imaging presented in Fig.~\ref{fig:cover}. From upper left to lower right, the panels display: CFHT u-band; Subaru HSC g, r, i, z, y bands; Spitzer/IRAC 3.6\,$\mu$m, 4.5\,$\mu$m, 5.8\,$\mu$m, 8.0\,$\mu$m; and Spitzer/MIPS 24\,$\mu$m. The upper row panels are 10"  in size, while the lower row panels are 40". The significant blending of the two nuclei and the central `beak' region in these lower-resolution images, particularly in the GALEX and Spitzer images, necessitates treating the system as a single composite source for SED modeling across the wavelength range from UV to IR.}
    \label{fig:sed_cutout}
\end{figure*}

\begin{figure}
    \centering
    \includegraphics[width=\linewidth]{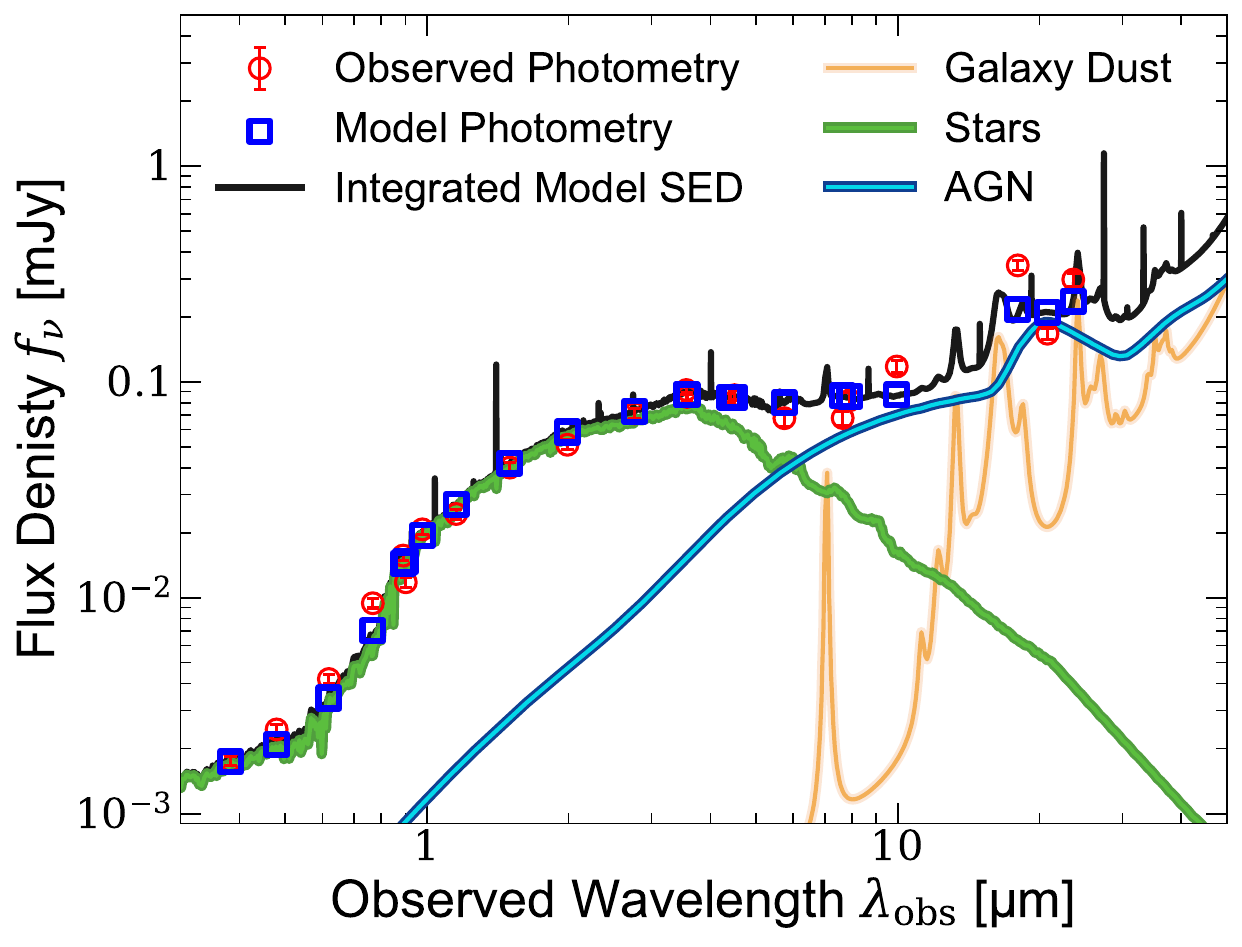}
    \caption{The spectral energy distribution (SED) modeling for the whole Cosmic Owl system. The integrated model SED is shown by the black curve. Individual components, including stars, galaxy dust, and AGN, are shown by green, orange, and blue curves, respectively. The observed photometry and model photometry are included in red circles (with uncertainties) and blue squares, respectively. The photometric data include GALEX/NUV, CFHT/u, Subaru-HSC/grizy, Spitzer/IRAC, MIPS, JWST NIRCam, and MIRI.}
    \label{fig:sed}
\end{figure}

The JWST imaging reveals that the target, dubbed the Cosmic Owl, exhibits rare twin ring structures.
Fig.~\ref{fig:cover} presents a pseudo-color composite of the Cosmic Owl using all JWST NIRCam wide-band filters, with individual band stamps shown in the lower panels.
The image vividly displays two symmetric ring structures, each associated with one of the merging galaxies.
The compact core of each galaxy forms an ``eye'' of the owl, while a central region of intense star formation, enhanced by younger stellar populations and nebular emission, appears in blue, resembling a ``beak'' between them.
These components are hereafter referred to as the NW eye, SE eye, and beak (see Fig.~\ref{fig:cover}).
Each ring has a diameter of approximately 1\arcsec, corresponding to about 8 kpc at $z=1.14$, comparable to the separation between the NW and SE eyes.
We note that neither of the two rings is empty, indicating that both rings feature a bluer outer ring and older inner stellar components.
We identify this system as twin collisional ring galaxies based on the criteria established by \citet{Appleton1996FCPh...16..111A}.
Collisional ring galaxies are characterized by a sharp, well-defined ring of young, blue stars surrounding a nucleus or offset bulge, often accompanied by radial ``spokes'' connecting the inner and outer regions.
These structures form when a companion galaxy passes through the disk of a target galaxy, triggering a radially propagating density wave.
The morphology of the Cosmic Owl perfectly matches this definition, specifically resembling the `RN' (Ring with Nucleus) subclass, where the nucleus remains prominent.
The symmetric nature of the rings suggests a rare, nearly head-on collision between two massive galaxies.

We perform spectral energy distribution (SED) modeling for the whole system, which combines the photometry of the two rings.
To measure photometry, first we convolve all JWST NIRCam images with a matching kernel to match the PSF of F444W band imaging (FWHM$\sim$0\farcs15).
The NIRCam PSFs are generated with the \textsc{WebbPSF} software \citep{Perrin2014SPIE.9143E..3XP} and the matching kernels are calculated using \textsc{photutils}.
Then, we conduct Kron photometry \citep{Kron1980ApJS...43..305K} on the PSF-matched images with a standard Kron radius $k=2.5$ for the entire system using \textsc{photutils}.
For the MIRI images, which have a large PSF (FWHM$\sim$0\farcs27--0\farcs67) compared to the galaxy size, we directly conduct Kron photometry on the mosaicked images without PSF matching.
In addition, we use the photometry from the COSMOS2020 catalog \citep{Weaver2022ApJS..258...11W}, including data taken by GALEX/NUV, CFHT/u, Subaru-HSC/grizy, Spitzer/IRAC, and MIPS.
These images utilized for the SED analysis are presented in Fig.~\ref{fig:sed_cutout}, in addition to the JWST imaged shown in Fig.~\ref{fig:cover}.
The SED modeling is carried out using the Code Investigating GALaxy Emission \citep[CIGALE;][]{Boquien2019A&A...622A.103B,Yang2022ApJ...927..192Y}.
For SED modeling, we assume a delayed-$\tau$ star formation history, where an optional late starburst is allowed in the last 20 Myr.
We use \citet{Bruzual2003MNRAS.344.1000B} stellar population synthesis models and an ionization parameter range of --1.0 to --3.0.
We adopt a modified \citet{Calzetti2000ApJ...533..682C} attenuation curve, which allows the variation of the power-law slope by $\pm0.3$.
We use the dust emission templates of \citet{Dale2014ApJ...784...83D} and AGN models from \citet{Stalevski2012MNRAS.420.2756S,Stalevski2016MNRAS.458.2288S}.
The redshift is fixed to $z=1.14$ when fitting, and a 5\% relative uncertainty floor of photometry is set.
The SED modeling results are shown in Fig.~\ref{fig:sed}.
The whole system exhibits a composite SED including massive stellar components with a total stellar mass of $M_\star=(3.2\pm0.4)\times10^{11}~M_\odot$ and AGN activity with 50\% AGN fraction.
We note that varying the assumed star formation history or the dust attenuation law introduces an additional systematic uncertainty of $\sim0.2-0.3$ dex in the stellar mass estimate.
We note that the SED modeling at $\sim20~\rm \mu m$ requires further improvement, which could result from the incomplete modeling of polycyclic aromatic hydrocarbon (PAH) emission and the $9.7~\rm \mu m$ silicate absorption features.

\subsection{Binary AGN}\label{ssec:AGN}

Multiple concurrent diagnostics indicate the presence of AGN in the Cosmic Owl: (1) the detection of broad hydrogen recombination lines (FWHM $> 2000$\,km\,s$^{-1}$), (2) luminous X-ray fluxes, (3) an extremely bright radio source with ultra-steep spectral index, (4) high-excitation nebular emission lines, and (5) SED modeling with significant contribution from the hot dust component.
We detail these robust indicators below.

The \target\ was identified as a bright X-ray source \texttt{cid\_1139} by the Chandra-COSMOS Legacy Survey \citep{Civano2016ApJ...819...62C, Marchesi2016ApJ...817...34M} with $f_{0.5-10\rm~keV}=(9.33\pm1.04)\times10^{-15}\rm~erg~s^{-1}~cm^{-2}$, corresponding to a X-ray luminosity of $L_{0.5-10\rm~keV}=(6.74\pm0.75)\times10^{43}\rm~erg~s^{-1}$.
This high luminosity alone strongly supports the presence of an AGN.
The spatial resolution ($\gtrsim1\arcsec$) of combined Chandra imaging \citep{Marchesi2016ApJ...817...34M} makes it hard to resolve individual components (see Appendix), though the system was identified as a candidate kpc-scale dual AGN by \citet{Li2024arXiv240514980L} based on X-ray selection.
Supporting this, the \target\ exhibits luminous radio continuum emission, detected as a $>10\sigma$ radio source ($1\sigma\sim0.14$~mJy) by the FIRST all-sky survey \citep{White1997ApJ...475..479W}.
As discussed in Sec.~\ref{ssec:radio}, this radio emission is synchrotron in origin.
Additionally, the Keck/DEIMOS spectrum (see Appendix) reveals highly ionized [\ion{Ne}{5}] emission lines, a unambiguous tracer of hard ionizing photons from an AGN \citep[\eg,][]{Gilli2010A&A...519A..92G,Izotov2012MNRAS.427.1229I,Mignoli2013A&A...556A..29M,Cleri2023ApJ...953...10C,Cleri2023ApJ...948..112C}.
Finally, the SED modeling described in Sec.~\ref{ssec:ring} indicates a significant contribution from hot dust and a 50\% AGN fraction.

\begin{figure*}
    \centering
    \includegraphics[width=\linewidth]{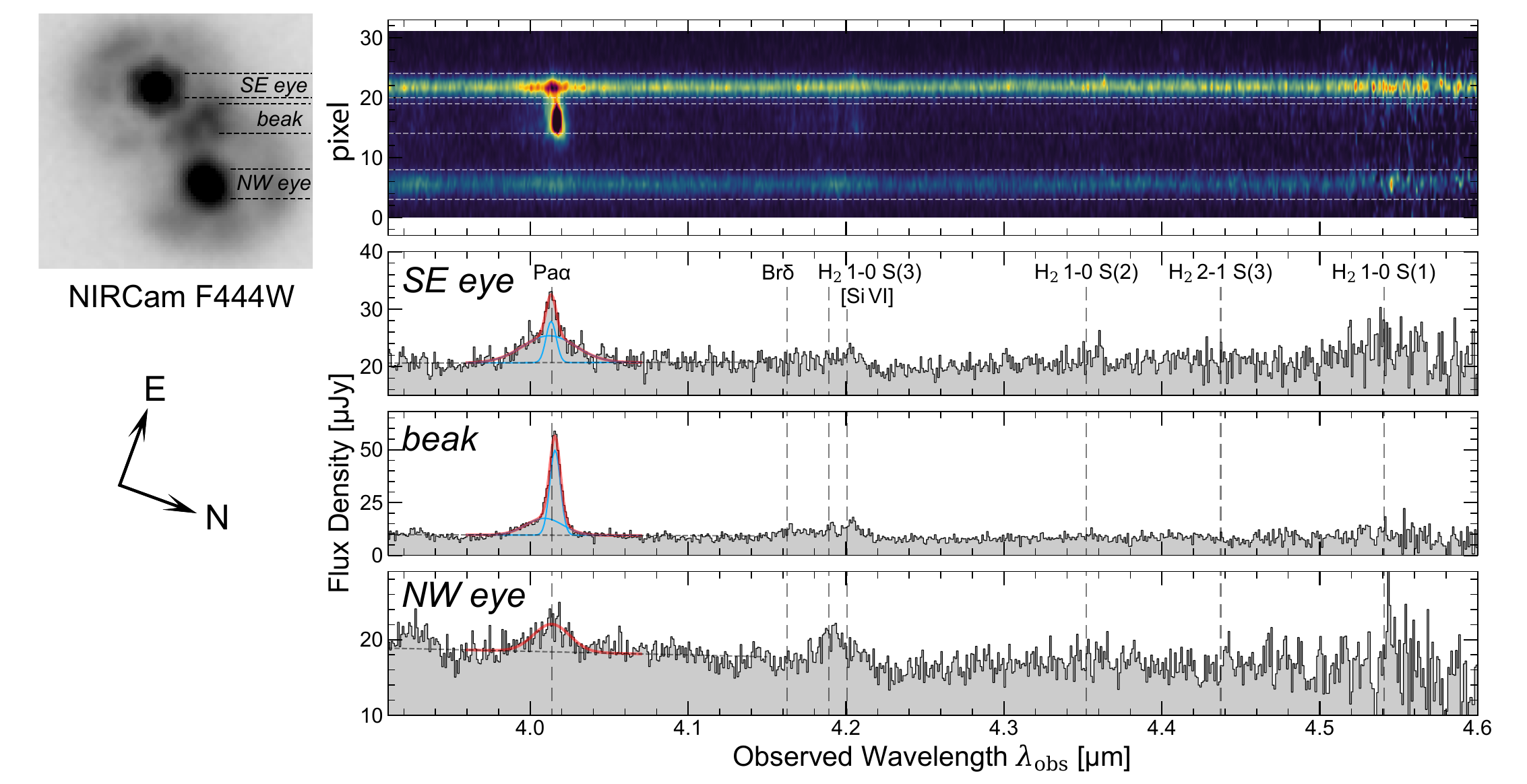}
    \caption{JWST NIRCam grism spectra for the Cosmic Owl from the COSMOS-3D program. The upper right panel shows the 2D spectrum with the extraction boxes overlaid as white dashed lines. The spatial axis has a pixel size of 0\farcs0629. The 1D spectra of the SE eye, beak, and NW eye are shown in the lower three panels. The wavelength of emission lines at $z=1.14$ is annotated as vertical dashed lines. The Gaussian models for the Pa$\alpha$ line are overlaid as red curves with blue curves showing individual Gaussian components, and the dashed gray lines denote the continuum models. The F444W image of the Cosmic Owl is shown in the upper left panel, rotated to match the grism position angle, with each component labeled. The JWST NIRCam grism spectra for the Cosmic Owl are available as the data behind the Figure in the online Journal.}
    \label{fig:c3d}
\end{figure*}

While the diagnostics above confirm AGN activity in the system, they are largely spatially unresolved.
To spatially disentangle the sources, we present grism spectra from the COSMOS-3D survey (Fig.~\ref{fig:c3d}), which directly confirm the existence of binary AGN in the SE eye and NW eye through broad hydrogen Paschen-$\alpha$ (Pa$\alpha$) emission lines.
We extract the 1D spectra of three components using boxcar apertures (lower panels of Fig.~\ref{fig:c3d}).
We model the continua with a linear function and the Pa$\alpha$ emission line with multiple Gaussian functions.
The Pa$\alpha$ line of the SE eye exhibits a broad component with FWHM$=2953\pm294~\mathrm{km\,s^{-1}}$ and a narrow component with FWHM$=567\pm78~\mathrm{km\,s^{-1}}$.
The line fluxes of the broad and narrow components are $f_\mathrm{SEb,Pa\alpha}=(3.68\pm0.44)\times10^{-17}~\mathrm{erg\,s^{-1}\,cm^{-2}}$ and $f_\mathrm{SEn,Pa\alpha}=(1.08\pm0.12)\times10^{-17}~\mathrm{erg\,s^{-1}\,cm^{-2}}$, respectively.
The Pa$\alpha$ line of the NW eye is modeled by an individual broad component with FWHM$=2018\pm250~\mathrm{km\,s^{-1}}$ and $f_\mathrm{NW,Pa\alpha}=(1.96\pm0.20)\times10^{-17}~\mathrm{erg\,s^{-1}\,cm^{-2}}$.
The broad Pa$\alpha$ components in both eyes display rather symmetric blue- and redshifted wings, typical of emission lines originating in the broad-line region (BLR) of AGN.

Using the broad Pa$\alpha$ line properties, we estimate the BH mass in the SE and NW eyes following the estimator presented by \citet{Kim2010ApJ...724..386K}:
\begin{equation}
\frac{M_\mathrm{BH}}{M_{\odot}}=10^{7.31}\left(\frac{L_{\mathrm{Pa} \alpha}}{L_{42}}\right)^{0.48}\left(\frac{\mathrm{FWHM}_{\mathrm{Pa} \alpha}}{\mathrm{FWHM}_{1000}}\right)^{1.68},
\end{equation}
where $L_{42} = 10^{42} ~\mathrm{erg} \mathrm{~s}^{-1}$ and $\mathrm{FWHM}_{1000}=10^3 \mathrm{~km} \mathrm{~s}^{-1}$.
The estimated BH masses are $M_\mathrm{BH,SE}=(6.7\pm2.9)\times10^{7}~M_\odot$ and $M_\mathrm{BH,NW}=(2.6\pm1.1)\times10^{7}~M_\odot$.
We note that the uncertainties in these BH masses are dominated by systematics ($\sim0.2$~dex) rather than measurement errors in the FWHM and line luminosity.
The projected distance between these two AGN is 1\farcs17 (9.62 kpc), showcasing a galaxy-scale binary AGN system.

\begin{figure*}
    \centering
    \includegraphics[width=\linewidth]{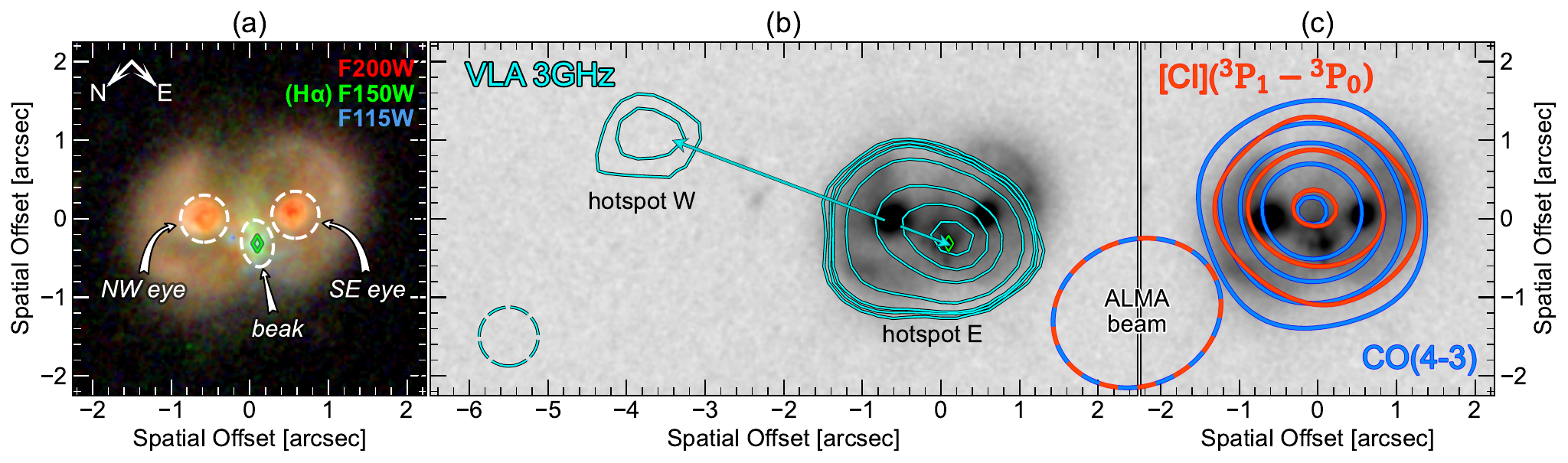}
    \includegraphics[width=\linewidth]{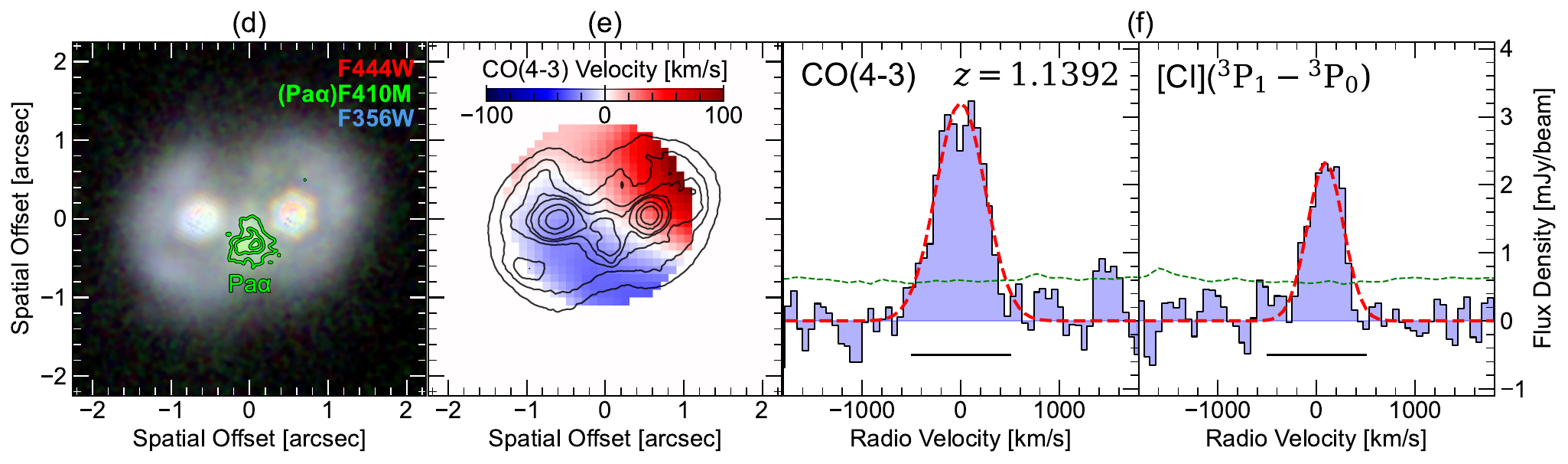}
    \caption{JWST, \textit{VLA}, and \textit{ALMA} observations for the \target.
    \textbf{(a)} Pseudo-color image of the Cosmic Owl with F115W, F150W, and F200W band images. The three components, including \textit{NW eye}, \textit{SE eye}, and the \textit{beak}, are annotated within. Note that the luminous H$\alpha$ emission line is in the F150W band, boosting the beak to be green. The two eyes are red due to the dust obscuration.
    \textbf{(b)} VLA 3GHz radio observation shown by cyan contours (3, 4, 5, 10, 30, 50, 70$\sigma$ with $1\sigma=2.4~\rm\mu Jy\,\mathrm{beam}^{-1}$), overlaid on the F200W image. The radio emission is spatially resolved into two hotspots, with a green diamond denoting the flux peak of hotspot E, coincident with the beak. The jet direction is indicated schematically by the cyan arrows.
    \textbf{(c)} ALMA observations of CO(4-3) (red lines) and \CI (blue lines), overlaid on the F200W image. The 3, 5, 7, 9, 11$\sigma$ contours are shown with $\sigma_\mathrm{CO}=0.166~\rm Jy/beam\cdot km/s$ and $\sigma_\mathrm{[CI]}=0.143~\rm Jy/beam\cdot km/s$. The CO and [CI] emission line maps are not spatially resolved; however, the emission peak is located between the two eyes and slightly ($\sim0.3$ arcsec) north of the beak's location.
    \textbf{(d)} Pseudo-color image made from the F356W, F410M, and F444W band images. The Pa$\alpha$ emission line is in the F410M band, boosting the beak to be green. The Pa$\alpha$ emission line map around the beak is overlaid as green contours with surface brightness of $(4,8,12)\times10^{-16}~\rm erg~s^{-1}~cm^{-2}~arcsec^{-2}$.
    \textbf{(e)} Mean CO(4-3) velocity field relative to the systemic redshift of the CO(4-3) emission ($z = 1.1392$). Only pixels brighter than 5$\sigma$ are included. The F277W image is overlaid as black contours.
    \textbf{(f)} Spectrum of the CO(4-3) and \CI lines from the \target. The velocity is relative to the systemic velocity of the CO(4-3) line. The dotted green line indicates the median 1$\sigma$ RMS uncertainty. Gaussian models fit to the data are shown in red dashed lines. The velocity range used to estimate the integrated line flux and mean velocity field is marked by the solid black bar.
    }
    \label{fig:almavla}
\end{figure*}

\subsection{Molecular Gas and Starburst Collisional Front}

We detect CO(4-3) and \CI emission in the Cosmic Owl with ALMA Band-6 observations.
The contours of the line intensity map of the CO(4-3) and \CI are shown in Fig.~\ref{fig:almavla}(c), and the peak extracted spectra are shown in Fig.~\ref {fig:almavla}(f).
We model the spectrum using a Gaussian profile, and the measured line fluxes of CO(4-3) and \CI are $I_{\rm CO(4-3)} = 2.01\pm0.19\rm~Jy~km~s^{-1}$ and $I_{\rm [CI]} = 1.05\pm0.18\rm~Jy~km~s^{-1}$, respectively.
The systemic redshift of the Cosmic Owl is $z=1.1392$, derived from the CO(4-3) line modeling.
The uncertainties in emission line fluxes are calculated by combining the fitting uncertainty and the noise in the spectrum.
Our measured line fluxes are consistent to within 1$\sigma$ with those reported in \citet{Valentino2020ApJ...890...24V}, which used different reduction procedures for the same data.
We note that we test the emission line modeling using a double Gaussian profile to estimate the total line flux, and we get a consistent result within 1$\sigma$ compared to the single Gaussian model.
The derived line luminosity is $L'_{\rm CO(4-3)}=(8.67\pm0.82)\times10^{9}~\mathrm{K\,km\,s^{-1}\,pc^{2}}$ and $L'_{\rm [CI]}=(3.97\pm0.68)\times10^{9}~\mathrm{K\,km\,s^{-1}\,pc^{2}}$.
Such luminous CO(4-3) and \CI lines trace a massive cold gas reservoir.
Based on the CO(4-3) emission, the molecular gas mass can be estimated as:
\begin{equation}
M_{\mathrm{mol}}\left[\mathrm{M}_{\odot}\right]=\alpha_{\mathrm{CO}}R_{41}^{-1} L_{\mathrm{CO}(4-3)}^{\prime},
\end{equation}
where the CO-to-$\rm H_2$ conversion factor $\alpha_\mathrm{CO}$ and the line brightness temperature ratio $R_{41}=L'_{\rm CO(4-3)}/L'_{\rm CO(1-0)}$ depend on the ISM properties \citep{Carilli2013ARA&A..51..105C}.
Typically, for starburst galaxies, $R_{41}=0.85$ is adopted, whereas for normal star-forming galaxies, the value of $R_{41}=0.17$ is used \citep{Bolatto2013ARA&A..51..207B}.
The CO-to-$\rm H_2$ conversion factor $\alpha_\mathrm{CO}$ typically ranges from $\alpha_\mathrm{CO}=0.8$ (starburst) to $3.6~\rm M_\odot/(\mathrm{K\,km\,s^{-1}\,pc^{2}})$ (normal star-forming).
Here, we adopt a canonical $\alpha_\mathrm{CO}=1.0~\rm M_\odot/(\mathrm{K\,km\,s^{-1}\,pc^{2}})$, a reference value chosen to facilitate comparison with other works.
These assumptions yield estimated molecular gas masses of $M_\mathrm{mol}=(1.0\pm0.1)\times10^{10}~\mathrm{M_\odot}$ (starburst case) and $M_\mathrm{mol}=(5.1\pm0.5)\times10^{10}~\mathrm{M_\odot}$ (normal star-forming case).
Note that these molecular gas mass estimates do not include a correction for helium.
While these derivations carry considerable uncertainty (up to $\sim1$ dex) due to the unconstrained gas conditions, the consensus suggests the presence of a substantial molecular gas reservoir of $\gtrsim10^{10}~\mathrm{M_\odot}$, likely concentrated in the ridge between the two galaxies.

\begin{figure*}[t]
    \centering
    \includegraphics[height=8cm]{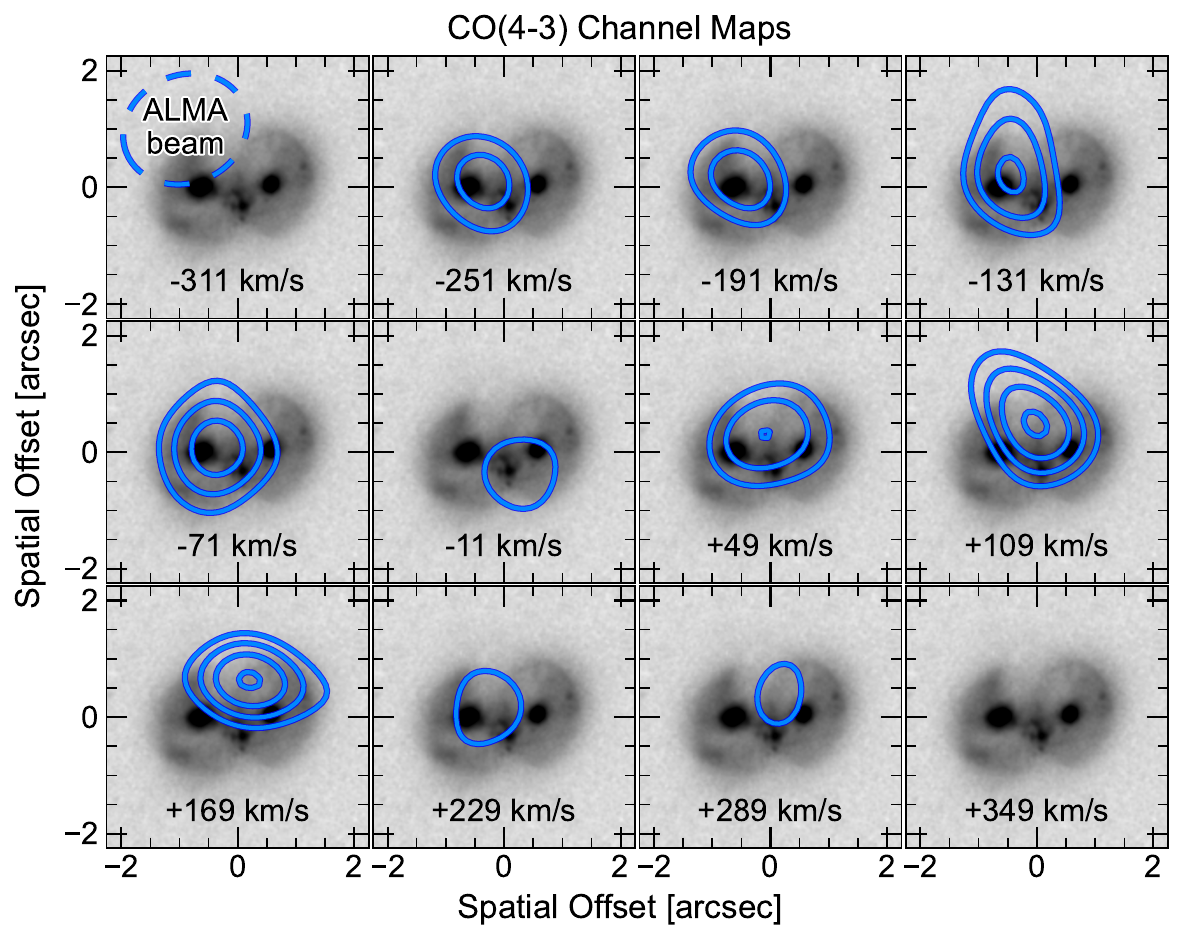}
    \includegraphics[height=8cm]{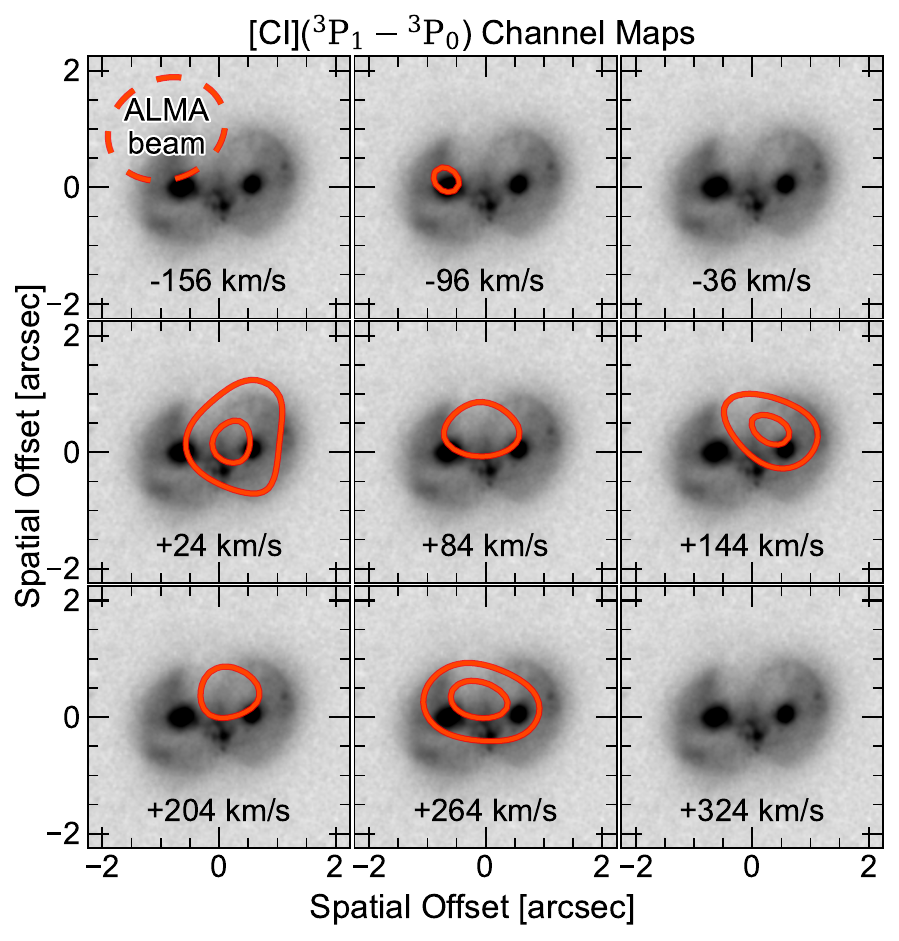}
    \caption{Velocity channel maps of the CO(4-3) (left) and \CI (right) emission in the Cosmic Owl. The channel width is $60~\rm km\,s^{-1}$, spanning a velocity range from approximately $-350~\rm km\,s^{-1}$ to $+350~\rm km\,s^{-1}$ relative to the systemic redshift. Channels with peak signal-to-noise ratios $<2$ are omitted. Contours start at 2$\sigma$ and increase in increments of 1$\sigma$, where $\sigma=0.62~\mathrm{mJy\,beam^{-1}}$ represents the rms noise per channel. The synthesized beam size is shown in the upper-left corner of the first panel. The emission is not spatially resolved, revealing a kinematic structure centered on the merging ridge, consistent with the gas reservoir fueling the starburst in the `beak' region. The ALMA CO(4-3) and \CI data cubes are available as the data behind the Figure in the online Journal.}
    \label{fig:alma_channel_map}
\end{figure*}

We find a clump located at the merging front, characterized by intense nebular emission lines.
The photometry in this region highlights strong UV and optical emission, indicative of recent star formation, likely triggered by the shock front from the galaxy collision and radio jets from the NW eye.
The F410M band covers the Hydrogen Pa$\alpha$ line, and we make a continuum-subtracted emission line image following the methods shown in \citet{Li2024ApJS..275...27L}.
We use the F444W band to estimate the continuum, assuming a flat $f_\nu$ continuum, which is suggested by the SED model (Fig.~\ref{fig:sed}). 
The Pa$\alpha$ emission line map is overlaid as green contours in Fig.~\ref{fig:almavla}, which is dominated around the beak region.
We measure the Pa$\alpha$ line flux in the beak region to be $(2.53\pm0.05)\times10^{-16}~\mathrm{erg\,s^{-1}\,cm^{-2}}$, which corresponds to a star formation rate of $84\pm2~\rm M_\odot\,yr^{-1}$, following the SFR calibration \citep{Kennicutt2012ARA&A..50..531K} and adopting a typical Case B recombination ratio between H$\alpha$ and Pa$\alpha$ to be 8.575 \citep{Osterbrock2006agna.book.....O}.
We note that this SFR is derived without correcting for attenuation by dust, so the SFR will be higher than this value, suggesting ongoing vigorous star formation in the beak region.
We also extract the NIRCam grism spectra of the beak region using a boxcar aperture and model the Pa$\alpha$ emission using multiple Gaussian components (Fig.~\ref{fig:c3d}).
The Pa$\alpha$ line of the beak is dominated by a narrow component with line flux of $(6.35\pm0.35)\times10^{-17}~\mathrm{erg\,s^{-1}\,cm^{-2}}$ and FWHM$=596\pm40~\mathrm{km\,s^{-1}}$.
A bluer broad component with a relative velocity of $-506\pm18~\mathrm{km\,s^{-1}}$ compared to the narrow component is revealed with line flux of $(4.34\pm0.92)\times10^{-17}~\mathrm{erg\,s^{-1}\,cm^{-2}}$ and FWHM$=2055\pm306~\mathrm{km\,s^{-1}}$.
We note that the blue wing of Pa$\alpha$ could be contaminated by \ion{He}{1} $\lambda1.868~\rm \mu m$, which is difficult to differentiate from the current data.
The Pa$\alpha$ rest-frame equivalent width in the extraction box is $\mathrm{EW_{0}(Pa\alpha) = 267\pm26~\AA}$.
We note that the total line flux measured from the grism spectra is lower than that derived from medium-band imaging, due to the limited width of the extraction box.
Together with the detection of a massive cold dense gas reservoir and the SFR derived from Pa$\alpha$ emission, confirms that a vigorous starburst is ongoing in the gas-rich beak region.

It is noteworthy that both the flux peaks of \CI and the CO line map are located between the two eyes (see Fig.~\ref{fig:almavla}c), close to the beak region, where we detect bright radio emission and strong nebular line emission.
The positional accuracy of the CO emission line peak \citep{Papadopoulos2008ApJ...684..845P} is estimated to be $\delta \theta_{\text {rms }} \sim 0.5\left\langle\Theta_{\text {beam }}\right\rangle(\mathrm{S} / \mathrm{N})^{-1} \sim 0\farcs1$, where the FWHM of the synthesized beam $\left\langle\Theta_{\text {beam }}\right\rangle\sim2\arcsec$ and the signal-to-noise ratio S/N$\sim$10 for the collapsed emission line image.
This is further confirmed by the channel maps of CO(4-3) and \CI emission lines (Fig.~\ref{fig:alma_channel_map}).
This means that the cold molecular and atomic gas, as traced by the \CI and CO lines, is at least partially distributed along the ridge between the two galaxies.
This suggests that this starburst clump is perhaps fueled by the cooling of gas shocked by the collision and the AGN jet, as shown below.
We show the mean CO(4-3) velocity field relative to the systemic redshift in Fig.~\ref{fig:almavla}e, which indicates a negative relative velocity in the northeast and a positive in the southwest, with a twisted distortion.
This presents tentative evidence of molecular gas funneled towards the beak \citep[\eg,][]{Wang2023ApJ...944..143W}.
In addition, the NIRCam grism spectra (Fig.~\ref{fig:c3d}) present a tentative detection of molecular Hydrogen line emission H$_2$ 1-0 S(3) in the beak region, which directly traces the warm molecular gas ($\gtrsim 10^3$~K) in such starburst sites \citep[e.g.,][]{Emonts2014A&A...572A..40E,Kristensen2023A&A...675A..86K}, consistent with the cold molecular gas detection.

\subsection{Bipolar Radio Jets}\label{ssec:radio}
The \target exhibits luminous radio continuum emission.
The deep VLA 3GHz radio imaging taken by the VLA-COSMOS project (VLA image hereafter) reveals the extended radio structure, which is shown in Fig~\ref{fig:almavla}b.
On a scale of about 6\arcsec or 50~kpc at the redshift of $z = 1.14$, the VLA image shows prominent bipolar radio structure along the east-west direction.
The VLA image has a 0\farcs75 diameter beam size, which cannot resolve the inner jets and only shows two extended radio lobes (hotspot W and hotspot E).
The hotspot W has a flux density of $S_{\rm W, 3~GHz} = 0.053 \pm 0.008$~mJy and the hotspot E has a flux density of $S_{\rm E, 3~GHz} = 0.430 \pm 0.008$~mJy measured using an aperture with radius of 1\farcs5.
The hotspot E peaks at the ``beak'' region with an extended radio structure connecting the NW eye and the beak.
The spatial alignment of the radio flux peak at hotspot E with the starburst region known as the ``beak" strongly suggests a physical link between the radio jet and the starburst activity.
The SMBH in the NW eye is likely the source of the relativistic radio jet, but a central radio core associated with the NW eye cannot be distinguished in the VLA image due to insufficient spatial resolution.
The radio emission peaking at the ``beak" region is likely produced by synchrotron radiation amplified by jet-induced shocks interacting with the interstellar medium \citep[\eg,][]{Lebowitz2023ApJ...951...73L}.
The projected distance between the flux peaks of eastern and western spots is 3\farcs8, or 31 kpc.
No galaxy is detected associated with hotspot W, even in the deep stacked image of all NIRCam bands.

\begin{figure}[h]
    \centering
    \includegraphics[width=\linewidth]{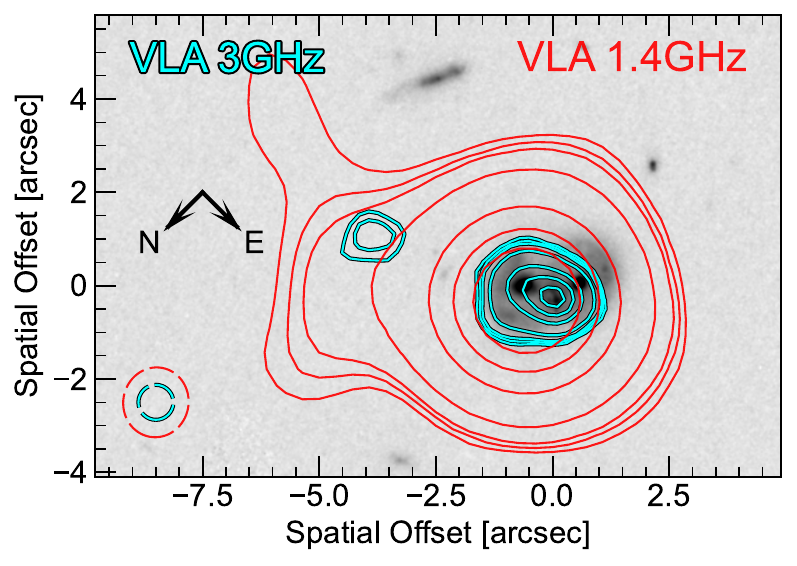}
    \caption{VLA 1.4\,GHz radio contours (red) at levels of 3, 4, 5, 10, 30, 50, and 70$\sigma$ ($1\sigma=10.8~\rm\mu Jy\,\mathrm{beam}^{-1}$) overlaid on the F200W image. For comparison, the higher-resolution VLA 3\,GHz imaging from Fig.~\ref{fig:almavla}b is also shown. While the 1.4\,GHz emission does not spatially resolve hotspot E, it partially resolves the whole radio source, revealing extended radio emission consistent with the opposite hotspot W observed at 3\,GHz with higher spatial resolution.}
    \label{fig:VLA1.4GHz}
\end{figure}

The radio imaging at 20~cm (1.4~GHz) can only partially resolve two hotspots with an angular resolution of $2\farcs5$, and only shows a tentative extended flux coinciding with the hotspot W revealed at 3 GHz (Fig.~\ref{fig:VLA1.4GHz}).
The total flux density at 20~cm is $1.66 \pm 0.17$~mJy measured using an aperture with radius of 5\arcsec.
The image at 90~cm has a spatial resolution of $8\arcsec\times6\arcsec$, therefore it cannot resolve any radio structures associated with the \target.
The peak flux density at 90~cm is $11.3\pm0.5$~mJy/beam.
These three band observations show an ultra-steep spectrum for the whole radio structure with an estimate of the spectral index ($S_\nu\propto \nu^{\alpha}$) to be $\alpha=-1.41\pm0.03$, consistent with synchrotron emission from high-$z$ radio AGN \citep{Miley2008A&ARv..15...67M}.

\begin{figure*}
    \centering
    \includegraphics[width=\linewidth]{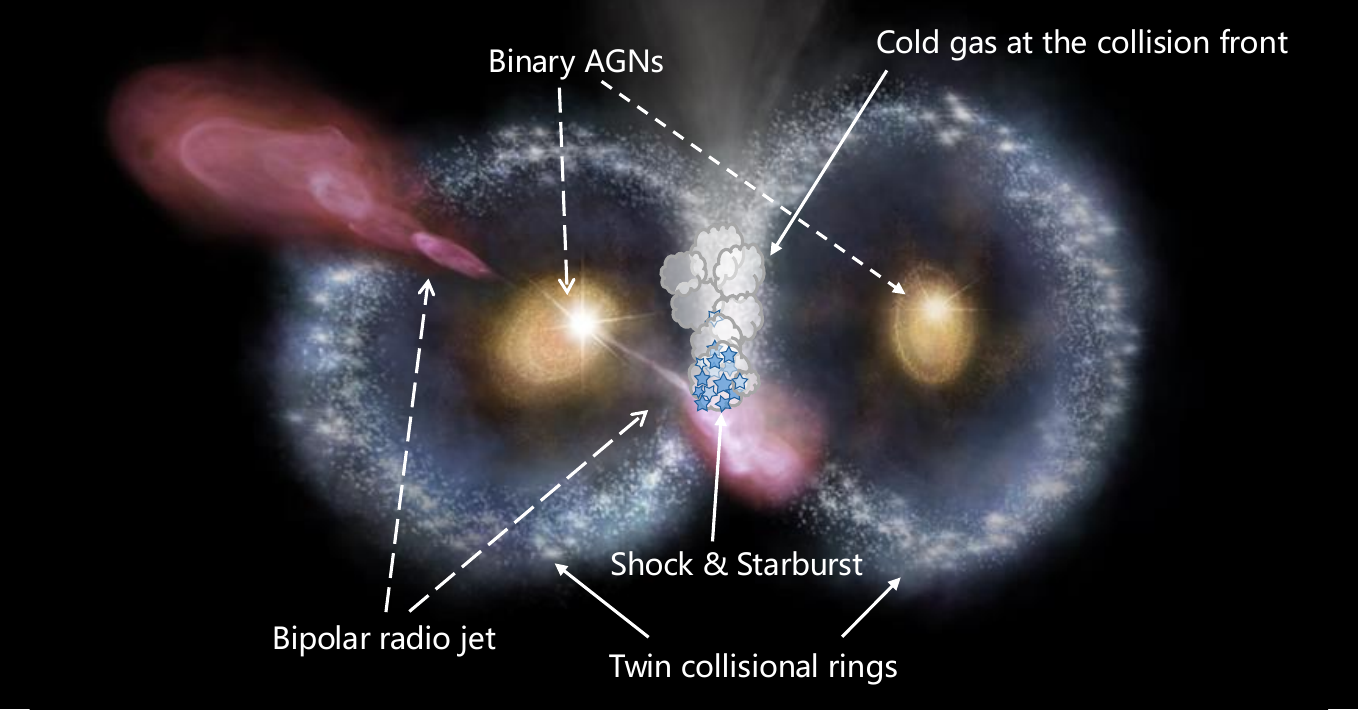}
    \caption{The schematic artistic view of the \target, consisting of twin collisional ring galaxies with binary AGN. One of the bipolar radio jets induces additional shocks in the collision front, where cold gas exists, and vigorous star formation occurs. This target showcases the intricate interplay of merger-driven ring formation, synchronized AGN activity, jet-induced feedback, and shock-induced star formation, offering a rare glimpse into the dynamical multifarious processes shaping galaxy evolution in the Universe.}
    \label{fig:sum}
\end{figure*}

\section{Discussion}\label{sec:discussion}
Multi-wavelength observations present a few concurrent physical phenomena within the \target, encompassing a head-on galaxy merger, twin collisional rings, a bipolar radio jet, binary AGN, and a starburst at the shocked collisional front (see Fig.~\ref{fig:sum} for a summary).
These phenomena mutually affect one another, collectively driving the evolution of this galaxy system, exemplifying a detailed view of galaxy mergers in the early Universe.
In this section, we explore the possible mechanisms and implications regarding these processes.

\subsection{Origin of the twin rings}

Collisional ring structures are thought to represent a rare, short-living phase that typically dissipates within a few hundred million years, as demonstrated by simulation studies \citep[e.g.,][]{Struck2010MNRAS.403.1516S}.
These rings form when a smaller or comparably sized galaxy passes through the disk of another, generating a gravitational shockwave.
This shockwave propagates outward, compressing gas into a dense ring, which triggers intense star formation and produces the bright, ring-like structures observed \citep{Appleton1996FCPh...16..111A}.
In the \target, both galaxies exhibit ring structures, and their symmetry suggests simultaneous formation, indicating that the two galaxies had similar masses and structures prior to a nearly central collision through their disks.
Assuming a relative velocity of $\sim250~\mathrm{km\,s^{-1}}$ (derived from the CO(4-3) line FWHM) and a projected separation of 9.62 kpc as the actual distance, the collision is estimated to have occurred approximately 38 Myr ago, aligning with the starburst timescale.
Using a ring radius of $\sim8$~kpc, the gravitational shockwave velocity is calculated to be approximately 208 km/s, consistent with the typical velocity of fast interstellar shock \citep[\eg,][]{Draine1993ARA&A..31..373D}.

While collisional ring galaxies are exceptionally rare, comprising only $\sim$0.01\% of galaxies in the local Universe \citep{Madore2009ApJS..181..572M}, the Cosmic Owl is unique even among this population due to its twin morphology.
Unlike typical systems, such as the Cartwheel Galaxy \citep{Amram1998A&A...330..881A} or Arp 147 \citep{Arp1966ApJS...14....1A}, where only one galaxy forms a ring, the Cosmic Owl involves two massive galaxies of comparable size, both of which develop rings simultaneously.
Furthermore, while radio detections in local ring galaxies are often associated with star formation, the detection of a radio-loud AGN jet interacting with the ring structure distinguishes the Cosmic Owl from the majority of known low-redshift examples.

To investigate the formation of the rare twin ring structure, we compare the Cosmic Owl with Arp 147, which is a pair of interacting ring galaxies formed by a collision between a spiral and an elliptical galaxy at $z = 0.03$ \citep{Gerber1992ApJ...399L..51G}.
In Arp 147, the collision produces a prominent, clumpy blue ring in the spiral galaxy, characterized by intense star formation within the past 40 Myr. The ring expands at approximately $225~\mathrm{km\,s^{-1}}$, with a collision timescale of 50 Myr \citep{Fogarty2011MNRAS.417..835F}.
The similarity in collision timescale and ring expansion velocity suggests that the Cosmic Owl likely formed through a comparable process to Arp 147, but with a more aligned collision angle and composed of twin galaxies with similar properties.
Further detailed merger simulations are required to explore the formation of such twin symmetrical rings of the Cosmic Owl.

Placing this system in the broader context of galaxy evolution, extensive wide-field studies using space telescopes have characterized the merger rate and its impact on galaxy structure at $z \gtrsim 1$ \citep[\eg,][]{Lackner2014AJ....148..137L, Duncan2019ApJ...876..110D}.
These works suggest that while major mergers are instrumental in building the massive end of the stellar mass function, identifying them in the short-lived collisional phase is rare.
The Cosmic Owl likely represents a brief but transformative stage in the assembly of a massive galaxy.
Given its substantial total stellar mass ($M_\ast \sim 3.2 \times 10^{11} M_\odot$) and the intense, shock-driven star formation consuming its gas reservoir, the predicted fate of this system is to evolve into a massive, gas-poor elliptical galaxy.
This aligns with the scenario where violent, gas-rich mergers at $z>1$ serve as the progenitors for the massive quiescent population observed at lower redshifts \citep[\eg,][]{Hopkins2006ApJS..163....1H}.

\subsection{Multifarious Processes in the Collisional Front}

The collisional front, especially the beak region in the \target, is shocked by the multifarious processes and exhibits intense star formation.
This starburst is likely triggered by the shock front generated during the head-on collision, which compresses gas and induces rapid cooling and collapse, as seen in other merger-driven systems \citep{Croft2006ApJ...647.1040C}.
The molecular and atomic gas, traced by CO(4-3) and \CI emission, peaks close to the beak region, suggesting that the cold gas reservoir is concentrated in the collisional front.
This gas distribution suggests a substantial molecular gas reservoir to sustain the observed starburst and potentially fuel further AGN activity. 
We note that, in addition to cooling, the cold gas may originate from tidal displacement caused by the merger.
The interaction between the gas and the bipolar radio jet, particularly at the hotspot E coincident with the beak, may further compress the gas, enhancing star formation through positive feedback mechanisms \citep{Fabian2012ARA&A..50..455F}.
All of these multifarious processes add layers of complexity.
However, it is precisely these intricate processes that shape the evolution of the galaxy ecosystem.
Future observations, such as high-resolution ALMA mapping of dense gas tracers, could clarify the jet’s impact on the molecular gas properties. In contrast, deeper high-resolution IFU spectroscopic data might reveal the ionization state and kinematics of the star-forming regions, providing a clearer picture of how these processes collectively shape the evolution of the Cosmic Owl.

The multifarious processes occurring in the Cosmic Owl, especially the enhanced star formation and molecular gas detected in the collisional front, bear resemblance to the properties of the nearby compact galaxy group Stephan’s Quintet ($z$\,=\,0.02, \eg, \citealt{Hickson1982ApJ...255..382H, Moles1998A&A...334..473M}).
For example, in Stephan’s Quintet, a ridge of molecular gas extends along a shock front that stretches midway between four interacting galaxies.
This ridge of molecular gas is detected in warm molecular H$_2$-emitting gas with {\it Spitzer} \citep{Appleton2006ApJ...639L..51A, Appleton2017ApJ...836...76A, Cluver2010ApJ...710..248C}, {\it Herschel} \citep{Appleton2013ApJ...777...66A}, and JWST \citep{Appleton2023ApJ...951..104A}, as well as in cold molecular gas through CO observations with {\it IRAM} \citep{Guillard2012ApJ...749..158G} and {\it ALMA} \citep{Appleton2023ApJ...951..104A,Emonts2025ApJ...978..111E}.
Along the extension of this ridge of molecular gas, a prominent region of star formation was detected in the North using {\it HST} \citep{Gallagher2001AJ....122..163G, Fedotov2011AJ....142...42F}, which has a counterpart also visible in radio continuum \citep{Xanthopoulos2004MNRAS.353.1117X}.
Shock models reveal that the formation of the molecular gas in the ridge is the result of dissipating mechanical energy along an elongated shock-front created by the interaction between the galaxies \citep{Guillard2009A&A...502..515G, Appleton2023ApJ...951..104A}.
One of the galaxies involved in this interaction has a Seyfert-2 AGN that contains a radio source \citep{Sulentic2001AJ....122.2993S, Aoki1999ApJ...521..565A, Xanthopoulos2004MNRAS.353.1117X, Pereira-Santaella2022A&A...665L..11P}.
The aligned ridge of cold gas ending in a region of prominent star formation in between the interacting galaxies resembles the morphology of the galaxies, cold gas, and enhanced star formation in the \target.
This adds to the credibility that molecular gas and ionized gas emission in the \target originates from gas cooling along a shock front between the merging galaxies.

Finally, we note that \citet{vanDokkum2025ApJ...988L...6V,vanDokkum2025ApJ...990L..48V} reported the independent discovery of this target, referring to it as the $\infty$ galaxy.
Their derivation of the system's fundamental properties is in excellent agreement with our findings.
Both studies determine a consistent redshift ($z=1.14$) and a comparable total stellar mass ($M_*\sim3\times10^{11}M_\odot$).
Furthermore, both works classify the morphology as a rare twin collisional ring system resulting from a nearly head-on collision.
Their follow-up investigation using JWST NIRSpec IFU observations \citep{vanDokkum2025ApJ...990L..48V} confirmed the presence of broad emission lines in both the SE and NW nuclei.
This aligns with our confirmation of binary AGNs based on the broad Pa$\alpha$ lines detected in the COSMOS-3D JWST NIRCam grism data.
The primary difference between the two interpretations concerns the physical nature of the central `beak' region located between the nuclei.
\citet{vanDokkum2025ApJ...988L...6V,vanDokkum2025ApJ...990L..48V} propose that the central source at the `beak' region hosts a third, active supermassive black hole that formed in situ via direct collapse, citing the centrally concentrated high-ionization emission lines and X-ray/radio detections.
Indeed, broad emission lines are detected at the `beak' region, which are also observed by the COSMOS-3D JWST grism spectra.
As discussed in \citet{vanDokkum2025ApJ...990L..48V}, this broad component can be evidence of BLR, but could also be a turbulent, dense gas in the outflow \citep[\eg,][]{Harrison2014MNRAS.441.3306H}.
In contrast, our multi-wavelength analysis suggests that the extreme conditions in the beak are driven by a starburst merging front compounded by a jet-ISM interaction.
As detailed in Sec.~3.4, the VLA images reveal an extended radio structure connecting the NW eye to the `beak' region and the radio lobe in the opposite direction, which supports the scenario of a bipolar jet originating from the NW AGN and striking the dense gas reservoir at the collision interface.
We argue that this jet-induced shock scenario can naturally account for the high-excitation lines (e.g., [\ion{Ne}{5}]), the synchrotron radio emission, and the vigorous star formation without requiring the presence of a third black hole.
Future high-resolution radio interferometry will be crucial to distinguish between a radio core (supporting a third direct-collapse black hole) and a hotspot (supporting the jet model) in the beak of the Cosmic Owl.

We also note that off-nuclear ultra-luminous X-ray sources (ULXs) are a ubiquitous feature of collisional ring galaxies, arising naturally from the intense, collision-induced star formation in the rings \citep[\eg,][]{Gao2003ApJ...596L.171G, Rappaport2010ApJ...721.1348R, Wolter2018ApJ...863...43W, Salvaggio2023MNRAS.522.1377S}.
In these systems, the formation of massive star clusters subsequently leads to the emergence of abundant massive X-ray binaries.
These ULXs can be extremely luminous with X-ray luminosity $L_\mathrm{X}$ reaching up to $\sim 10^{42}\rm \, erg\,s^{-1}$ for an individual ULX \citep[\eg,][]{Farrell2009Natur.460...73F}, and they are typically believed to be powered by stellar-mass or intermediate-mass black holes accreting at super-Eddington rates \citep{Kaaret2017ARA&A..55..303K, King2023NewAR..9601672K}, rather than supermassive black holes.
In the case of the Cosmic Owl, which resembles collisional ring galaxies, the vigorous starburst at the collision front makes the presence of abundant ULXs statistically probable.
These sources likely contribute to the X-ray emission, potentially explaining the X-ray flux peak and the broad lines detected at the `beak' region.
Further studies are required to disentangle the multifarious physical processes occurring in the Cosmic Owl.

\section{Summary}\label{sec:summary}
We present a comprehensive, multi-wavelength analysis of the Cosmic Owl, a galaxy merger at $z = 1.14$, identified in the COSMOS-CANDELS field.
Our investigation, leveraging public archival data including VLA, ALMA, and JWST (PRIMER, COSMOS-Web, and COSMOS-3D), reveals a unique interacting system composed of twin AGN-host collisional ring galaxies with a starburst merging front.
The main results are summarized as follows:
\begin{enumerate}
    \item The Cosmic Owl consists of two interacting galaxies that have formed nearly identical collisional ring structures, each with a diameter of approximately 8 kpc. The stellar mass of the whole system is $M_\star=(3.2\pm0.4)\times10^{11}~M_\odot$.  The symmetry of the rings suggests a head-on collision origin between two galaxies of similar mass and structure.
    \item We confirm the presence of an active galactic nucleus (AGN) in each of the merging galaxies. This is evidenced by broad Pa$\alpha$ emission lines ($>2000~\mathrm{km\,s^{-1}}$) detected in the JWST/NIRCam grism spectra, as well as supported by X-ray ($L_{0.5-10\rm~keV}=(6.74\pm0.75)\times10^{43}\rm~erg~s^{-1}$) detection, luminous radio continuum emission ($S_\mathrm{3\,GHz}=1.66\pm0.17~\mathrm{mJy}$), and detection of high-excitation emission line [\ion{Ne}{5}]$\lambda\lambda3347,3427$. This showcases a galactic-scale binary AGN with a projected separation of $9.62$~kpc. The BH masses for the SE eye and NW eye are measured to be $M_\mathrm{BH,SE}=(6.7\pm2.9)\times10^{7}~M_\odot$ and $M_\mathrm{BH,NW}=(2.6\pm1.1)\times10^{7}~M_\odot$, respectively.
    \item A bipolar radio jet, originating from the AGN in the NW eye, extends to the beak region of the Cosmic Owl and induces additional shocks in the collision front between the two galaxies. This region hosts a vigorous starburst ($\rm SFR = 84\pm2~M_\odot\,yr^{-1}$), characterized by intense nebular line emission (e.g., $f_\mathrm{Pa\alpha}=(2.53\pm0.05)\times10^{-16}~\mathrm{erg\,s^{-1}\,cm^{-2}}$), and a massive cold molecular gas reservoir ($\gtrsim10^{10}~\mathrm{M_\odot}$).
    The confluence of a galaxy collision and an AGN jet at the merging front provides a compelling case for shock-induced star formation. These shocks appear to be compressing the interstellar medium, leading to the rapid and efficient conversion of gas into stars.
\end{enumerate}

The Cosmic Owl system serves as an exceptional laboratory for studying the multifaceted processes that drive galaxy evolution.
The simultaneous occurrence of a head-on merger, twin-ring formation, dual AGN activity, and a jet-triggered starburst offers a detailed snapshot of the mechanisms that assemble stellar mass and grow supermassive black holes in the early universe. 
The rare twin-ring structure calls for dedicated numerical simulations to constrain the precise initial conditions of the merger.
Furthermore, this system highlights the pivotal role of shocks in triggering intense starbursts.
These shocks, induced by galaxy collisions or relativistic jets, can compress gas at the collision front and drive rapid, efficient conversion of molecular gas into stars.
At higher redshift, where mergers are more frequent, as suggested by recent JWST surveys \citep[\eg,][]{Duan2025MNRAS.540..774D}, such collision-triggered starbursts may represent a previously under-appreciated channel for boosting early cosmic star formation.

\begin{acknowledgments}
The authors gratefully acknowledge artist Hua Fu for creating the schematic artistic view of the Cosmic Owl (Fig.~\ref{fig:sum}).
The spectacular jets shown in Fig.~\ref{fig:sum} are adapted from the image of radio galaxy \href{https://esahubble.org/images/opo1247a/}{Hercules A} (Credit: NASA, ESA, S. Baum and C. O'Dea (RIT), R. Perley and W. Cotton (NRAO/AUI/NSF), and the Hubble Heritage Team (STScI/AURA)).
The authors thank Olivier Ilbert, Mara Salvato, and Ali Ahmad Khostovan for their help with Keck/DEIMOS data.
The authors thank Shude Mao and Zechang Sun for helpful discussion.
This work is supported by National Key R\&D Program of China (grant no. 2023YFA1605600) and Tsinghua University Initiative Scientific Research Program.
TT is supported by JSPS KAKENHI Grant Number JP25KJ0750 and the Forefront Physics and Mathematics Program to Drive Transformation (FoPM), a World-leading Innovative Graduate Study (WINGS) Program at the University of Tokyo.
FV acknowledges support from the Independent Research Fund Denmark (DFF) under grant 3120-00043B and the Cosmic Dawn Center (DAWN), which is funded by the Danish National Research Foundation under grant DNRF140.
This work is based in part on observations made with the NASA/ESA/CSA James Webb Space Telescope. The data were obtained from the Mikulski Archive for Space Telescopes at the Space Telescope Science Institute, which is operated by the Association of Universities for Research in Astronomy, Inc., under NASA contract NAS 5-03127 for JWST. These observations are associated with program \#1727, \#1837, and \#5893.
The authors acknowledge the COSMOS-Web team led by coPIs (J. Kartaltepe and C. Casey), PRIMER team led by PI (J. Dunlop), and COSMOS-3D team led by CoPIs (K. Kakiichi, X. Fan, F. Wang, E. Egami, J. Lyu, and J. Yang) for developing their observing program with a zero-exclusive-access period.
All the JWST raw data used in this paper can be found in MAST: \dataset[10.17909/e6db-b169]{http://dx.doi.org/10.17909/e6db-b169}.
This paper makes use of the following ALMA data: ADS/JAO.ALMA\#2016.1.01040.S. ALMA is a partnership of ESO (representing its member states), NSF (USA) and NINS (Japan), together with NRC (Canada), MOST and ASIAA (Taiwan), and KASI (Republic of Korea), in cooperation with the Republic of Chile. The Joint ALMA Observatory is operated by ESO, AUI/NRAO and NAOJ.
The National Radio Astronomy Observatory is a facility of the National Science Foundation operated under cooperative agreement by Associated Universities, Inc.
\end{acknowledgments}





%
\facilities{JWST(NIRCam, MIRI), JVLA, ALMA, Keck:II(DEIMOS), CXO}



\appendix

\section{Archival data with AGN evidence}

\begin{figure*}[h]
    \centering
    \includegraphics[height=7.2cm]{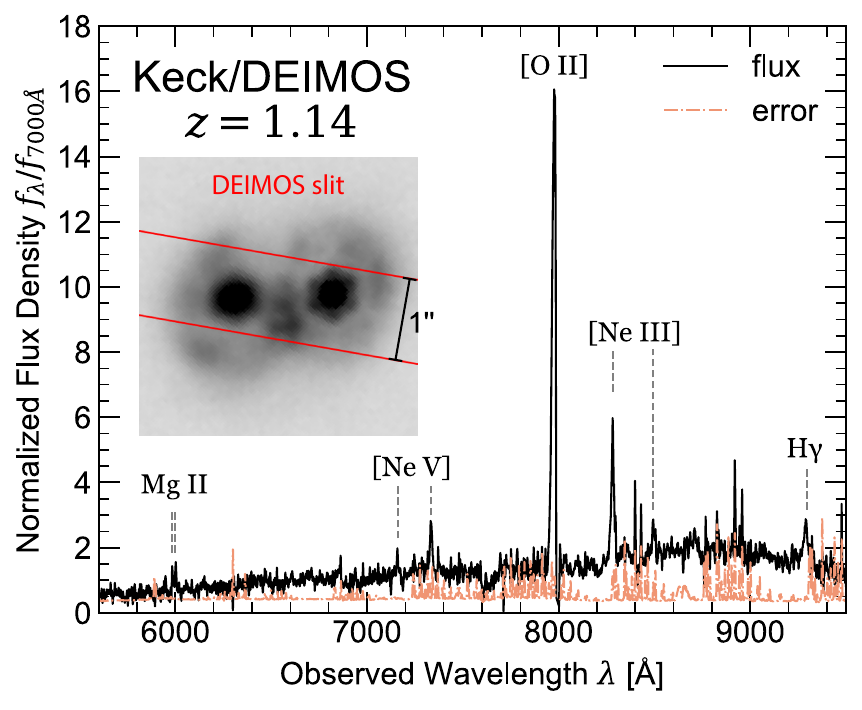}
    \includegraphics[height=7.2cm]{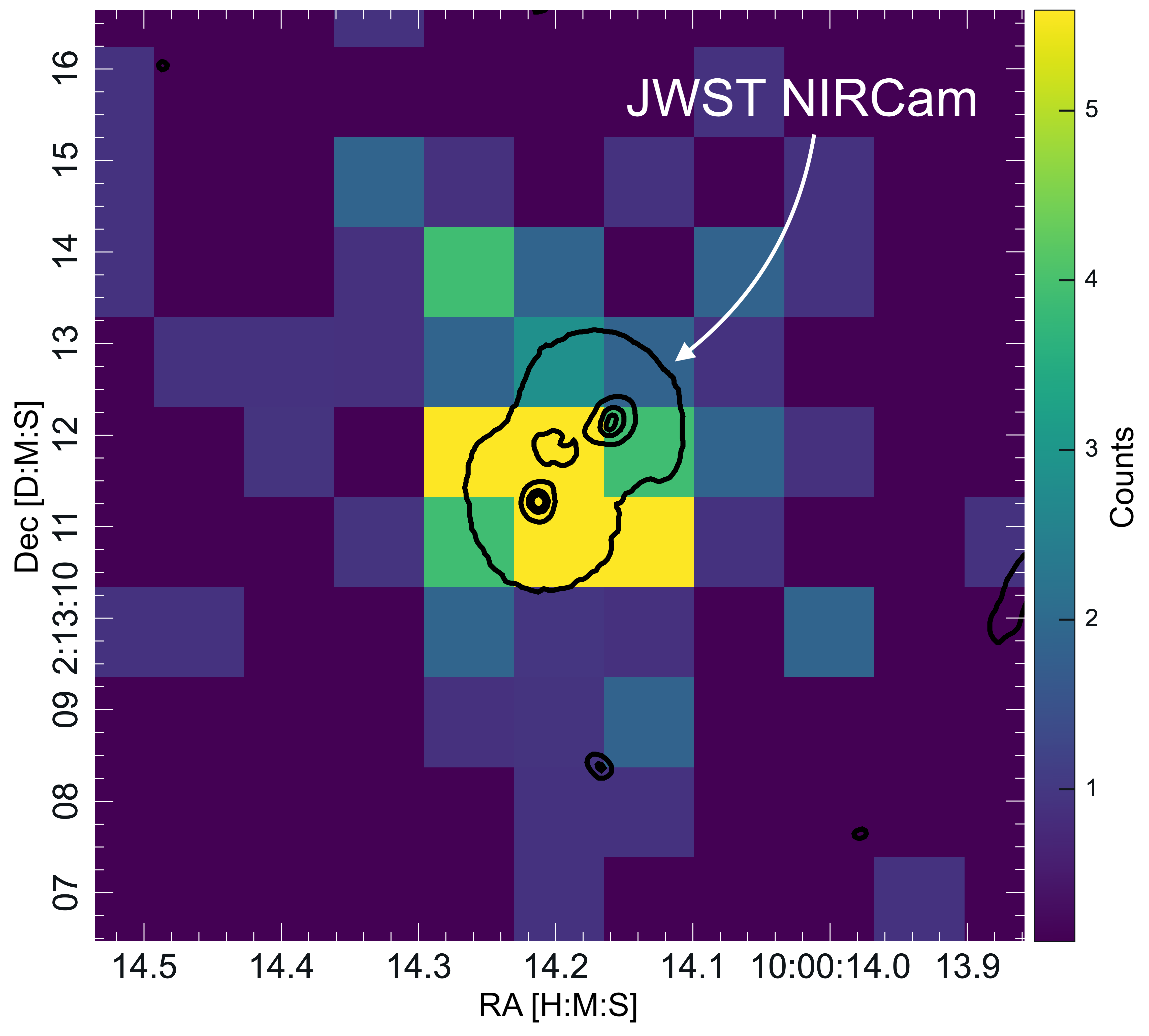}
    \caption{Complementary evidence from archival data to support the presence of AGN in the Cosmic Owl.
    Left: Keck/DEIMOS spectrum of the \target. The black lines denote the flux density with error overlaid as orange dotted lines. A series of detected emission lines, including [\ion{O}{2}], [\ion{Ne}{3}], [\ion{Ne}{5}], H$\gamma$, and \ion{Mg}{2}, have been annotated. The spectrum is normalized at the observed wavelength of 7000\,\AA. The DEIMOS slit is shown in the inner panel, overlaid on the JWST F444W image with the same rotation as Fig.~\ref{fig:cover}. Right: Chandra 0.5-7\,keV X-ray imaging of the Cosmic Owl with JWST NIRCam imaging overlaid as black contours. In this figure, north is at the top and east is to the left.}
    \label{fig:deimos_spec}
    \label{fig:xray}
\end{figure*}

The \target has been observed through multi-slit spectroscopy with the Deep Imaging Multi-Object Spectrograph (DEIMOS) on the Keck II telescope, collected as DEIMOS10K survey \citep{Hasinger2018ApJ...858...77H}.
We directly use the archival product of the extracted 1D spectrum, which is \href{https://irsa.ipac.caltech.edu/data/COSMOS/spectra/deimos/spec1d_tbl/spec1d.CSN1a.051.SPIRE-cc-29266.tbl}{publicly available}.
The target ID of the \target is L607340 in DEIMOS10K survey and the slit position angle is 125$^\circ$.
The DEIMOS spectrum (shown in Fig.~\ref{fig:deimos_spec}) reveals a series of emission lines, including [\ion{O}{2}], [\ion{Ne}{3}], [\ion{Ne}{5}], H$\gamma$, and \ion{Mg}{2}.
The significant detection of the very high-excitation emission line [\ion{Ne}{5}] $\lambda\lambda3347,3427$ indicates an extreme ionization and confirms the existence of AGN or strong shocks \citep[\eg,][]{Gilli2010A&A...519A..92G,Izotov2012MNRAS.427.1229I,Mignoli2013A&A...556A..29M,Cleri2023ApJ...953...10C,Cleri2023ApJ...948..112C}, because it requires $h\nu > 97.12$~eV photons to make the highly ionized \ion{Ne}{5} emission line.

The Cosmic Owl was detected as a bright X-ray source by Chandra imaging (see Fig.~\ref{fig:xray}).
The exposure time is 182.63\,ks and the 0.5-7 keV count rate is $(5.94\pm0.66)\times10^{-4}$ counts~s$^{-1}$.
We refer to \citealt{Civano2016ApJ...819...62C, Marchesi2016ApJ...817...34M} for detailed description of the Chandra data.
We note that this combined X-ray imaging in the COSMOS field has a point-spread function (defined as 50\% of the encircled energy fraction in 0.5–7 keV) with a size larger than $1\arcsec$, which cannot resolve the individual components in the Cosmic Owl.

\bibliography{main}{}

@ARTICLE{Lackner2014AJ....148..137L,
       author = {{Lackner}, C.~N. and {Silverman}, J.~D. and {Salvato}, M. and {Kampczyk}, P. and {Kartaltepe}, J.~S. and {Sanders}, D. and {Capak}, P. and {Civano}, F. and {Halliday}, C. and {Ilbert}, O. and {Jahnke}, K. and {Koekemoer}, A.~M. and {Lee}, N. and {Le F{\`e}vre}, O. and {Liu}, C.~T. and {Scoville}, N. and {Sheth}, K. and {Toft}, S.},
        title = "{Late-Stage Galaxy Mergers in Cosmos to z {\ensuremath{\sim}} 1}",
      journal = {AJ},
         year = 2014,
        month = dec,
       volume = {148},
       number = {6},
          eid = {137},
        pages = {137},
          doi = {10.1088/0004-6256/148/6/137},
archivePrefix = {arXiv},
       eprint = {1406.2327},
 primaryClass = {astro-ph.GA},
       adsurl = {https://ui.adsabs.harvard.edu/abs/2014AJ....148..137L}
}

@ARTICLE{Farrell2009Natur.460...73F,
       author = {{Farrell}, Sean A. and {Webb}, Natalie A. and {Barret}, Didier and {Godet}, Olivier and {Rodrigues}, Joana M.},
        title = "{An intermediate-mass black hole of over 500 solar masses in the galaxy ESO243-49}",
      journal = {Natur},
         year = 2009,
        month = jul,
       volume = {460},
       number = {7251},
        pages = {73-75},
          doi = {10.1038/nature08083},
archivePrefix = {arXiv},
       eprint = {1001.0567},
 primaryClass = {astro-ph.HE},
       adsurl = {https://ui.adsabs.harvard.edu/abs/2009Natur.460...73F}
}

@ARTICLE{Rappaport2010ApJ...721.1348R,
       author = {{Rappaport}, S. and {Levine}, A. and {Pooley}, D. and {Steinhorn}, B.},
        title = "{Ultraluminous X-ray Sources in Arp 147}",
      journal = {ApJ},
         year = 2010,
        month = oct,
       volume = {721},
       number = {2},
        pages = {1348-1355},
          doi = {10.1088/0004-637X/721/2/1348},
archivePrefix = {arXiv},
       eprint = {1007.3271},
 primaryClass = {astro-ph.GA},
       adsurl = {https://ui.adsabs.harvard.edu/abs/2010ApJ...721.1348R}
}

@ARTICLE{Kaaret2017ARA&A..55..303K,
       author = {{Kaaret}, Philip and {Feng}, Hua and {Roberts}, Timothy P.},
        title = "{Ultraluminous X-Ray Sources}",
      journal = {ARA\&A},
         year = 2017,
        month = aug,
       volume = {55},
       number = {1},
        pages = {303-341},
          doi = {10.1146/annurev-astro-091916-055259},
archivePrefix = {arXiv},
       eprint = {1703.10728},
 primaryClass = {astro-ph.HE},
       adsurl = {https://ui.adsabs.harvard.edu/abs/2017ARA&A..55..303K}
}

@ARTICLE{King2023NewAR..9601672K,
       author = {{King}, Andrew and {Lasota}, Jean-Pierre and {Middleton}, Matthew},
        title = "{Ultraluminous X-ray sources}",
      journal = {NewAR},
         year = 2023,
        month = jun,
       volume = {96},
          eid = {101672},
        pages = {101672},
          doi = {10.1016/j.newar.2022.101672},
archivePrefix = {arXiv},
       eprint = {2302.10605},
 primaryClass = {astro-ph.HE},
       adsurl = {https://ui.adsabs.harvard.edu/abs/2023NewAR..9601672K}
}

@ARTICLE{Salvaggio2023MNRAS.522.1377S,
       author = {{Salvaggio}, Chiara and {Wolter}, A. and {Belfiore}, A. and {Colpi}, M.},
        title = "{The largest bright ULX population in a galaxy: X-ray variability and luminosity function in the Cartwheel ring galaxy}",
      journal = {MNRAS},
         year = 2023,
        month = jun,
       volume = {522},
       number = {1},
        pages = {1377-1393},
          doi = {10.1093/mnras/stad943},
archivePrefix = {arXiv},
       eprint = {2303.15502},
 primaryClass = {astro-ph.HE},
       adsurl = {https://ui.adsabs.harvard.edu/abs/2023MNRAS.522.1377S}
}

@ARTICLE{Wolter2018ApJ...863...43W,
       author = {{Wolter}, Anna and {Fruscione}, Antonella and {Mapelli}, Michela},
        title = "{The X-Ray Luminosity Function of Ultraluminous X-Ray Sources in Collisional Ring Galaxies}",
      journal = {ApJ},
         year = 2018,
        month = aug,
       volume = {863},
       number = {1},
          eid = {43},
        pages = {43},
          doi = {10.3847/1538-4357/aacb34},
archivePrefix = {arXiv},
       eprint = {1806.02746},
 primaryClass = {astro-ph.HE},
       adsurl = {https://ui.adsabs.harvard.edu/abs/2018ApJ...863...43W}
}

@ARTICLE{Gao2003ApJ...596L.171G,
       author = {{Gao}, Yu and {Wang}, Q. Daniel and {Appleton}, P.~N. and {Lucas}, Ray A.},
        title = "{Nonnuclear Hyper/Ultraluminous X-Ray Sources in the Starbursting Cartwheel Ring Galaxy}",
      journal = {ApJL},
         year = 2003,
        month = oct,
       volume = {596},
       number = {2},
        pages = {L171-L174},
          doi = {10.1086/379598},
archivePrefix = {arXiv},
       eprint = {astro-ph/0309253},
 primaryClass = {astro-ph},
       adsurl = {https://ui.adsabs.harvard.edu/abs/2003ApJ...596L.171G}
}

@ARTICLE{Harrison2014MNRAS.441.3306H,
       author = {{Harrison}, C.~M. and {Alexander}, D.~M. and {Mullaney}, J.~R. and {Swinbank}, A.~M.},
        title = "{Kiloparsec-scale outflows are prevalent among luminous AGN: outflows and feedback in the context of the overall AGN population}",
      journal = {MNRAS},
         year = 2014,
        month = jul,
       volume = {441},
       number = {4},
        pages = {3306-3347},
          doi = {10.1093/mnras/stu515},
archivePrefix = {arXiv},
       eprint = {1403.3086},
 primaryClass = {astro-ph.GA},
       adsurl = {https://ui.adsabs.harvard.edu/abs/2014MNRAS.441.3306H}
}

@ARTICLE{Lebowitz2023ApJ...951...73L,
       author = {{Lebowitz}, Sophie and {Emonts}, Bjorn and {Terndrup}, Donald M. and {Burchett}, Joseph N. and {Prochaska}, J. Xavier and {Drouart}, Guillaume and {Villar-Mart{\'\i}n}, Montserrat and {Lehnert}, Matthew and {De Breuck}, Carlos and {Vernet}, Jo{\"e}l and {Alatalo}, Katherine},
        title = "{The Dragonfly Galaxy. III. Jet Brightening of a High-redshift Radio Source Caught in a Violent Merger of Disk Galaxies}",
      journal = {ApJ},
         year = 2023,
        month = jul,
       volume = {951},
       number = {1},
          eid = {73},
        pages = {73},
          doi = {10.3847/1538-4357/acd3ed},
archivePrefix = {arXiv},
       eprint = {2305.05564},
 primaryClass = {astro-ph.GA},
       adsurl = {https://ui.adsabs.harvard.edu/abs/2023ApJ...951...73L}
}

@ARTICLE{Bolatto2013ARA&A..51..207B,
       author = {{Bolatto}, Alberto D. and {Wolfire}, Mark and {Leroy}, Adam K.},
        title = "{The CO-to-H$_{2}$ Conversion Factor}",
      journal = {ARA\&A},
         year = 2013,
        month = aug,
       volume = {51},
       number = {1},
        pages = {207-268},
          doi = {10.1146/annurev-astro-082812-140944},
archivePrefix = {arXiv},
       eprint = {1301.3498},
 primaryClass = {astro-ph.GA},
       adsurl = {https://ui.adsabs.harvard.edu/abs/2013ARA&A..51..207B}
}

@ARTICLE{Amram1998A&A...330..881A,
       author = {{Amram}, P. and {Mendes de Oliveira}, C. and {Boulesteix}, J. and {Balkowski}, C.},
        title = "{The H{\ensuremath{\alpha}} kinematic of the Cartwheel galaxy}",
      journal = {A\&A},
         year = 1998,
        month = feb,
       volume = {330},
        pages = {881-893},
       adsurl = {https://ui.adsabs.harvard.edu/abs/1998A&A...330..881A}
}

@ARTICLE{vanDokkum2025ApJ...990L..48V,
       author = {{van Dokkum}, Pieter and {Brammer}, Gabriel and {Jennings}, Connor and {Pasha}, Imad and {Baggen}, Josephine F.~W.},
        title = "{Further Evidence for a Direct-collapse Origin of the Supermassive Black Hole at the Center of the {\ensuremath{\infty}} Galaxy}",
      journal = {ApJL},
         year = 2025,
        month = sep,
       volume = {990},
       number = {2},
          eid = {L48},
        pages = {L48},
          doi = {10.3847/2041-8213/adfb50},
archivePrefix = {arXiv},
       eprint = {2506.15619},
 primaryClass = {astro-ph.GA},
       adsurl = {https://ui.adsabs.harvard.edu/abs/2025ApJ...990L..48V}
}

@ARTICLE{vanDokkum2025ApJ...988L...6V,
       author = {{van Dokkum}, Pieter and {Brammer}, Gabriel and {Baggen}, Josephine F.~W. and {Keim}, Michael A. and {Natarajan}, Priyamvada and {Pasha}, Imad},
        title = "{The {\ensuremath{\infty}} Galaxy: A Candidate Direct-collapse Supermassive Black Hole between Two Massive, Ringed Nuclei}",
      journal = {ApJL},
         year = 2025,
        month = jul,
       volume = {988},
       number = {1},
          eid = {L6},
        pages = {L6},
          doi = {10.3847/2041-8213/addcfe},
archivePrefix = {arXiv},
       eprint = {2506.15618},
 primaryClass = {astro-ph.GA},
       adsurl = {https://ui.adsabs.harvard.edu/abs/2025ApJ...988L...6V}
}

@software{bushouse_2025_15178003,
  author       = {Bushouse, Howard and
                  Eisenhamer, Jonathan and
                  Dencheva, Nadia and
                  Davies, James and
                  Greenfield, Perry and
                  Morrison, Jane and
                  Hodge, Phil and
                  Simon, Bernie and
                  Grumm, David and
                  Droettboom, Michael and
                  Slavich, Edward and
                  Sosey, Megan and
                  Pauly, Tyler and
                  Miller, Todd and
                  Jedrzejewski, Robert and
                  Hack, Warren and
                  Davis, David and
                  Crawford, Steven and
                  Law, David and
                  Gordon, Karl and
                  Regan, Michael and
                  Cara, Mihai and
                  MacDonald, Ken and
                  Bradley, Larry and
                  Shanahan, Clare and
                  Jamieson, William and
                  Teodoro, Mairan and
                  Williams, Thomas and
                  Pena-Guerrero, Maria and
                  Graham, Brett and
                  Molter, Edward and
                  Brandt, Timothy and
                  Hayes, Christian and
                  Cooper, Rachel and
                  Clarke, Melanie and
                  Filippazzo, Joseph},
  title        = {JWST Calibration Pipeline},
  month        = apr,
  year         = 2025,
  publisher    = {Zenodo},
  version      = {1.18.0},
  doi          = {10.5281/zenodo.15178003},
  url          = {https://doi.org/10.5281/zenodo.15178003},
  swhid        = {swh:1:dir:31be401baa8e9a98a870d7063fa2203dd60cd4e9
                   ;origin=https://doi.org/10.5281/zenodo.6984365;vis
                   it=swh:1:snp:23c4ec4fdb7b9d73ff763f2ea44b5a4b06419
                   f79;anchor=swh:1:rel:4cb75fcb301902eb92daadde71d6c
                   e10beb371c4;path=spacetelescope-jwst-ddf32b8
                  },
}

@misc{sun_2024_14052875,
  author       = {Sun, Fengwu},
  title        = {nircam\_grism},
  month        = jul,
  year         = 2024,
  publisher    = {Zenodo},
  version      = {3.0},
  doi          = {10.5281/zenodo.14052875},
  url          = {https://doi.org/10.5281/zenodo.14052875},
}

@ARTICLE{Aihara2022PASJ...74..247A,
       author = {{Aihara}, Hiroaki and {AlSayyad}, Yusra and {Ando}, Makoto and {Armstrong}, Robert and {Bosch}, James and {Egami}, Eiichi and {Furusawa}, Hisanori and {Furusawa}, Junko and {Harasawa}, Sumiko and {Harikane}, Yuichi and {Hsieh}, Bau-Ching and {Ikeda}, Hiroyuki and {Ito}, Kei and {Iwata}, Ikuru and {Kodama}, Tadayuki and {Koike}, Michitaro and {Kokubo}, Mitsuru and {Komiyama}, Yutaka and {Li}, Xiangchong and {Liang}, Yongming and {Lin}, Yen-Ting and {Lupton}, Robert H. and {Lust}, Nate B. and {MacArthur}, Lauren A. and {Mawatari}, Ken and {Mineo}, Sogo and {Miyatake}, Hironao and {Miyazaki}, Satoshi and {More}, Surhud and {Morishima}, Takahiro and {Murayama}, Hitoshi and {Nakajima}, Kimihiko and {Nakata}, Fumiaki and {Nishizawa}, Atsushi J. and {Oguri}, Masamune and {Okabe}, Nobuhiro and {Okura}, Yuki and {Ono}, Yoshiaki and {Osato}, Ken and {Ouchi}, Masami and {Pan}, Yen-Chen and {Plazas Malag{\'o}n}, Andr{\'e}s A. and {Price}, Paul A. and {Reed}, Sophie L. and {Rykoff}, Eli S. and {Shibuya}, Takatoshi and {Simunovic}, Mirko and {Strauss}, Michael A. and {Sugimori}, Kanako and {Suto}, Yasushi and {Suzuki}, Nao and {Takada}, Masahiro and {Takagi}, Yuhei and {Takata}, Tadafumi and {Takita}, Satoshi and {Tanaka}, Masayuki and {Tang}, Shenli and {Taranu}, Dan S. and {Terai}, Tsuyoshi and {Toba}, Yoshiki and {Turner}, Edwin L. and {Uchiyama}, Hisakazu and {Vijarnwannaluk}, Bovornpratch and {Waters}, Christopher Z. and {Yamada}, Yoshihiko and {Yamamoto}, Naoaki and {Yamashita}, Takuji},
        title = "{Third data release of the Hyper Suprime-Cam Subaru Strategic Program}",
      journal = {PASJ},
         year = 2022,
        month = apr,
       volume = {74},
       number = {2},
        pages = {247-272},
          doi = {10.1093/pasj/psab122},
archivePrefix = {arXiv},
       eprint = {2108.13045},
 primaryClass = {astro-ph.IM},
       adsurl = {https://ui.adsabs.harvard.edu/abs/2022PASJ...74..247A}
}

@ARTICLE{Aoki1999ApJ...521..565A,
       author = {{Aoki}, Kentaro and {Kosugi}, George and {Wilson}, Andrew S. and {Yoshida}, Michitoshi},
        title = "{The Radio Emission of the Seyfert Galaxy NGC 7319}",
      journal = {ApJ},
         year = 1999,
        month = aug,
       volume = {521},
       number = {2},
        pages = {565-571},
          doi = {10.1086/307559},
archivePrefix = {arXiv},
       eprint = {astro-ph/9812365},
 primaryClass = {astro-ph},
       adsurl = {https://ui.adsabs.harvard.edu/abs/1999ApJ...521..565A}
}

@ARTICLE{Appleton1996FCPh...16..111A,
       author = {{Appleton}, P.~N. and {Struck-Marcell}, C.},
        title = "{Collisional Ring Galaxies}",
      journal = {FCPh},
         year = 1996,
        month = jan,
       volume = {16},
        pages = {111-220},
       adsurl = {https://ui.adsabs.harvard.edu/abs/1996FCPh...16..111A}
}

@ARTICLE{Appleton2006ApJ...639L..51A,
       author = {{Appleton}, P.~N. and {Xu}, Kevin C. and {Reach}, William and {Dopita}, Michael A. and {Gao}, Y. and {Lu}, N. and {Popescu}, C.~C. and {Sulentic}, J.~W. and {Tuffs}, R.~J. and {Yun}, M.~S.},
        title = "{Powerful High-Velocity Dispersion Molecular Hydrogen Associated with an Intergalactic Shock Wave in Stephan's Quintet}",
      journal = {ApJL},
         year = 2006,
        month = mar,
       volume = {639},
       number = {2},
        pages = {L51-L54},
          doi = {10.1086/502646},
archivePrefix = {arXiv},
       eprint = {astro-ph/0602554},
 primaryClass = {astro-ph},
       adsurl = {https://ui.adsabs.harvard.edu/abs/2006ApJ...639L..51A}
}

@ARTICLE{Appleton2013ApJ...777...66A,
       author = {{Appleton}, P.~N. and {Guillard}, P. and {Boulanger}, F. and {Cluver}, M.~E. and {Ogle}, P. and {Falgarone}, E. and {Pineau des For{\^e}ts}, G. and {O'Sullivan}, E. and {Duc}, P. -A. and {Gallagher}, S. and {Gao}, Y. and {Jarrett}, T. and {Konstantopoulos}, I. and {Lisenfeld}, U. and {Lord}, S. and {Lu}, N. and {Peterson}, B.~W. and {Struck}, C. and {Sturm}, E. and {Tuffs}, R. and {Valchanov}, I. and {van der Werf}, P. and {Xu}, K.~C.},
        title = "{Shock-enhanced C$^{+}$ Emission and the Detection of H$_{2}$O from the Stephan's Quintet Group-wide Shock Using Herschel}",
      journal = {ApJ},
         year = 2013,
        month = nov,
       volume = {777},
       number = {1},
          eid = {66},
        pages = {66},
          doi = {10.1088/0004-637X/777/1/66},
archivePrefix = {arXiv},
       eprint = {1309.1525},
 primaryClass = {astro-ph.CO},
       adsurl = {https://ui.adsabs.harvard.edu/abs/2013ApJ...777...66A}
}

@ARTICLE{Appleton2017ApJ...836...76A,
       author = {{Appleton}, P.~N. and {Guillard}, P. and {Togi}, A. and {Alatalo}, K. and {Boulanger}, F. and {Cluver}, M. and {Pineau des For{\^e}ts}, G. and {Lisenfeld}, U. and {Ogle}, P. and {Xu}, C.~K.},
        title = "{Powerful H$_{2}$ Line Cooling in Stephan{\textquoteright}s Quintet. II. Group-wide Gas and Shock Modeling of the Warm H$_{2}$ and a Comparison with [C II] 157.7 {\ensuremath{\mu}}m Emission and Kinematics}",
      journal = {ApJ},
         year = 2017,
        month = feb,
       volume = {836},
       number = {1},
          eid = {76},
        pages = {76},
          doi = {10.3847/1538-4357/836/1/76},
archivePrefix = {arXiv},
       eprint = {1701.03226},
 primaryClass = {astro-ph.GA},
       adsurl = {https://ui.adsabs.harvard.edu/abs/2017ApJ...836...76A}
}

@ARTICLE{Appleton2023ApJ...951..104A,
       author = {{Appleton}, P.~N. and {Guillard}, P. and {Emonts}, Bjorn and {Boulanger}, Francois and {Togi}, Aditya and {Reach}, William T. and {Alatalo}, Katherine and {Cluver}, M. and {Diaz Santos}, T. and {Duc}, P. -A. and {Gallagher}, S. and {Ogle}, P. and {O'Sullivan}, E. and {Voggel}, K. and {Xu}, C.~K.},
        title = "{Multiphase Gas Interactions on Subarcsec Scales in the Shocked Intergalactic Medium of Stephan's Quintet with JWST and ALMA}",
      journal = {ApJ},
         year = 2023,
        month = jul,
       volume = {951},
       number = {2},
          eid = {104},
        pages = {104},
          doi = {10.3847/1538-4357/accc2a},
archivePrefix = {arXiv},
       eprint = {2301.02928},
 primaryClass = {astro-ph.GA},
       adsurl = {https://ui.adsabs.harvard.edu/abs/2023ApJ...951..104A}
}

@ARTICLE{Arp1966ApJS...14....1A,
       author = {{Arp}, Halton},
        title = "{Atlas of Peculiar Galaxies}",
      journal = {The Astrophysical Journal Supplement Series},
         year = 1966,
        month = nov,
       volume = {14},
        pages = {1},
          doi = {10.1086/190147},
       adsurl = {https://ui.adsabs.harvard.edu/abs/1966ApJS...14....1A}
}

@ARTICLE{Boquien2019A&A...622A.103B,
       author = {{Boquien}, M. and {Burgarella}, D. and {Roehlly}, Y. and {Buat}, V. and {Ciesla}, L. and {Corre}, D. and {Inoue}, A.~K. and {Salas}, H.},
        title = "{CIGALE: a python Code Investigating GALaxy Emission}",
      journal = {A\&A},
         year = 2019,
        month = feb,
       volume = {622},
          eid = {A103},
        pages = {A103},
          doi = {10.1051/0004-6361/201834156},
archivePrefix = {arXiv},
       eprint = {1811.03094},
 primaryClass = {astro-ph.GA},
       adsurl = {https://ui.adsabs.harvard.edu/abs/2019A&A...622A.103B}
}

@ARTICLE{Bruzual2003MNRAS.344.1000B,
       author = {{Bruzual}, G. and {Charlot}, S.},
        title = "{Stellar population synthesis at the resolution of 2003}",
      journal = {MNRAS},
         year = 2003,
        month = oct,
       volume = {344},
       number = {4},
        pages = {1000-1028},
          doi = {10.1046/j.1365-8711.2003.06897.x},
archivePrefix = {arXiv},
       eprint = {astro-ph/0309134},
 primaryClass = {astro-ph},
       adsurl = {https://ui.adsabs.harvard.edu/abs/2003MNRAS.344.1000B}
}

@ARTICLE{CASA2022PASP..134k4501C,
       author = {{CASA Team} and {Bean}, Ben and {Bhatnagar}, Sanjay and {Castro}, Sandra and {Donovan Meyer}, Jennifer and {Emonts}, Bjorn and {Garcia}, Enrique and {Garwood}, Robert and {Golap}, Kumar and {Gonzalez Villalba}, Justo and {Harris}, Pamela and {Hayashi}, Yohei and {Hoskins}, Josh and {Hsieh}, Mingyu and {Jagannathan}, Preshanth and {Kawasaki}, Wataru and {Keimpema}, Aard and {Kettenis}, Mark and {Lopez}, Jorge and {Marvil}, Joshua and {Masters}, Joseph and {McNichols}, Andrew and {Mehringer}, David and {Miel}, Renaud and {Moellenbrock}, George and {Montesino}, Federico and {Nakazato}, Takeshi and {Ott}, Juergen and {Petry}, Dirk and {Pokorny}, Martin and {Raba}, Ryan and {Rau}, Urvashi and {Schiebel}, Darrell and {Schweighart}, Neal and {Sekhar}, Srikrishna and {Shimada}, Kazuhiko and {Small}, Des and {Steeb}, Jan-Willem and {Sugimoto}, Kanako and {Suoranta}, Ville and {Tsutsumi}, Takahiro and {van Bemmel}, Ilse M. and {Verkouter}, Marjolein and {Wells}, Akeem and {Xiong}, Wei and {Szomoru}, Arpad and {Griffith}, Morgan and {Glendenning}, Brian and {Kern}, Jeff},
        title = "{CASA, the Common Astronomy Software Applications for Radio Astronomy}",
      journal = {\pasp},
         year = 2022,
        month = nov,
       volume = {134},
       number = {1041},
          eid = {114501},
        pages = {114501},
          doi = {10.1088/1538-3873/ac9642},
archivePrefix = {arXiv},
       eprint = {2210.02276},
 primaryClass = {astro-ph.IM},
       adsurl = {https://ui.adsabs.harvard.edu/abs/2022PASP..134k4501C}
}

@ARTICLE{Calzetti2000ApJ...533..682C,
       author = {{Calzetti}, Daniela and {Armus}, Lee and {Bohlin}, Ralph C. and {Kinney}, Anne L. and {Koornneef}, Jan and {Storchi-Bergmann}, Thaisa},
        title = "{The Dust Content and Opacity of Actively Star-forming Galaxies}",
      journal = {ApJ},
         year = 2000,
        month = apr,
       volume = {533},
       number = {2},
        pages = {682-695},
          doi = {10.1086/308692},
archivePrefix = {arXiv},
       eprint = {astro-ph/9911459},
 primaryClass = {astro-ph},
       adsurl = {https://ui.adsabs.harvard.edu/abs/2000ApJ...533..682C}
}

@ARTICLE{Carilli2013ARA&A..51..105C,
       author = {{Carilli}, C.~L. and {Walter}, F.},
        title = "{Cool Gas in High-Redshift Galaxies}",
      journal = {ARA\&A},
         year = 2013,
        month = aug,
       volume = {51},
       number = {1},
        pages = {105-161},
          doi = {10.1146/annurev-astro-082812-140953},
archivePrefix = {arXiv},
       eprint = {1301.0371},
 primaryClass = {astro-ph.CO},
       adsurl = {https://ui.adsabs.harvard.edu/abs/2013ARA&A..51..105C}
}

@ARTICLE{Casey2023ApJ...954...31C,
       author = {{Casey}, Caitlin M. and {Kartaltepe}, Jeyhan S. and {Drakos}, Nicole E. and {Franco}, Maximilien and {Harish}, Santosh and {Paquereau}, Louise and {Ilbert}, Olivier and {Rose}, Caitlin and {Cox}, Isabella G. and {Nightingale}, James W. and {Robertson}, Brant E. and {Silverman}, John D. and {Koekemoer}, Anton M. and {Massey}, Richard and {McCracken}, Henry Joy and {Rhodes}, Jason and {Akins}, Hollis B. and {Allen}, Natalie and {Amvrosiadis}, Aristeidis and {Arango-Toro}, Rafael C. and {Bagley}, Micaela B. and {Bongiorno}, Angela and {Capak}, Peter L. and {Champagne}, Jaclyn B. and {Chartab}, Nima and {Ch{\'a}vez Ortiz}, {\'O}scar A. and {Chworowsky}, Katherine and {Cooke}, Kevin C. and {Cooper}, Olivia R. and {Darvish}, Behnam and {Ding}, Xuheng and {Faisst}, Andreas L. and {Finkelstein}, Steven L. and {Fujimoto}, Seiji and {Gentile}, Fabrizio and {Gillman}, Steven and {Gould}, Katriona M.~L. and {Gozaliasl}, Ghassem and {Hayward}, Christopher C. and {He}, Qiuhan and {Hemmati}, Shoubaneh and {Hirschmann}, Michaela and {Jahnke}, Knud and {Jin}, Shuowen and {Khostovan}, Ali Ahmad and {Kokorev}, Vasily and {Lambrides}, Erini and {Laigle}, Clotilde and {Larson}, Rebecca L. and {Leung}, Gene C.~K. and {Liu}, Daizhong and {Liaudat}, Tobias and {Long}, Arianna S. and {Magdis}, Georgios and {Mahler}, Guillaume and {Mainieri}, Vincenzo and {Manning}, Sinclaire M. and {Maraston}, Claudia and {Martin}, Crystal L. and {McCleary}, Jacqueline E. and {McKinney}, Jed and {McPartland}, Conor J.~R. and {Mobasher}, Bahram and {Pattnaik}, Rohan and {Renzini}, Alvio and {Rich}, R. Michael and {Sanders}, David B. and {Sattari}, Zahra and {Scognamiglio}, Diana and {Scoville}, Nick and {Sheth}, Kartik and {Shuntov}, Marko and {Sparre}, Martin and {Suzuki}, Tomoko L. and {Talia}, Margherita and {Toft}, Sune and {Trakhtenbrot}, Benny and {Urry}, C. Megan and {Valentino}, Francesco and {Vanderhoof}, Brittany N. and {Vardoulaki}, Eleni and {Weaver}, John R. and {Whitaker}, Katherine E. and {Wilkins}, Stephen M. and {Yang}, Lilan and {Zavala}, Jorge A.},
        title = "{COSMOS-Web: An Overview of the JWST Cosmic Origins Survey}",
      journal = {ApJ},
         year = 2023,
        month = sep,
       volume = {954},
       number = {1},
          eid = {31},
        pages = {31},
          doi = {10.3847/1538-4357/acc2bc},
archivePrefix = {arXiv},
       eprint = {2211.07865},
 primaryClass = {astro-ph.GA},
       adsurl = {https://ui.adsabs.harvard.edu/abs/2023ApJ...954...31C}
}

@ARTICLE{Cielo2018MNRAS.477.1336C,
       author = {{Cielo}, Salvatore and {Bieri}, Rebekka and {Volonteri}, Marta and {Wagner}, Alexander Y. and {Dubois}, Yohan},
        title = "{AGN feedback compared: jets versus radiation}",
      journal = {MNRAS},
         year = 2018,
        month = jun,
       volume = {477},
       number = {1},
        pages = {1336-1355},
          doi = {10.1093/mnras/sty708},
archivePrefix = {arXiv},
       eprint = {1712.03955},
 primaryClass = {astro-ph.GA},
       adsurl = {https://ui.adsabs.harvard.edu/abs/2018MNRAS.477.1336C}
}

@ARTICLE{Civano2016ApJ...819...62C,
       author = {{Civano}, F. and {Marchesi}, S. and {Comastri}, A. and {Urry}, M.~C. and {Elvis}, M. and {Cappelluti}, N. and {Puccetti}, S. and {Brusa}, M. and {Zamorani}, G. and {Hasinger}, G. and {Aldcroft}, T. and {Alexander}, D.~M. and {Allevato}, V. and {Brunner}, H. and {Capak}, P. and {Finoguenov}, A. and {Fiore}, F. and {Fruscione}, A. and {Gilli}, R. and {Glotfelty}, K. and {Griffiths}, R.~E. and {Hao}, H. and {Harrison}, F.~A. and {Jahnke}, K. and {Kartaltepe}, J. and {Karim}, A. and {LaMassa}, S.~M. and {Lanzuisi}, G. and {Miyaji}, T. and {Ranalli}, P. and {Salvato}, M. and {Sargent}, M. and {Scoville}, N.~J. and {Schawinski}, K. and {Schinnerer}, E. and {Silverman}, J. and {Smolcic}, V. and {Stern}, D. and {Toft}, S. and {Trakhtenbrot}, B. and {Treister}, E. and {Vignali}, C.},
        title = "{The Chandra Cosmos Legacy Survey: Overview and Point Source Catalog}",
      journal = {ApJ},
         year = 2016,
        month = mar,
       volume = {819},
       number = {1},
          eid = {62},
        pages = {62},
          doi = {10.3847/0004-637X/819/1/62},
archivePrefix = {arXiv},
       eprint = {1601.00941},
 primaryClass = {astro-ph.GA},
       adsurl = {https://ui.adsabs.harvard.edu/abs/2016ApJ...819...62C}
}

@ARTICLE{Clauwens2018MNRAS.478.3994C,
       author = {{Clauwens}, Bart and {Schaye}, Joop and {Franx}, Marijn and {Bower}, Richard G.},
        title = "{The three phases of galaxy formation}",
      journal = {MNRAS},
         year = 2018,
        month = aug,
       volume = {478},
       number = {3},
        pages = {3994-4009},
          doi = {10.1093/mnras/sty1229},
archivePrefix = {arXiv},
       eprint = {1711.00030},
 primaryClass = {astro-ph.GA},
       adsurl = {https://ui.adsabs.harvard.edu/abs/2018MNRAS.478.3994C}
}

@ARTICLE{Cleri2023ApJ...948..112C,
       author = {{Cleri}, Nikko J. and {Yang}, Guang and {Papovich}, Casey and {Trump}, Jonathan R. and {Backhaus}, Bren E. and {Estrada-Carpenter}, Vicente and {Finkelstein}, Steven L. and {Giavalisco}, Mauro and {Hutchison}, Taylor A. and {Ji}, Zhiyuan and {Jung}, Intae and {Matharu}, Jasleen and {Momcheva}, Ivelina and {Olivier}, Grace M. and {Simons}, Raymond and {Weiner}, Benjamin},
        title = "{CLEAR: High-ionization [Ne V] {\ensuremath{\lambda}}3426 Emission-line Galaxies at 1.4 < z < 2.3}",
      journal = {ApJ},
         year = 2023,
        month = may,
       volume = {948},
       number = {2},
          eid = {112},
        pages = {112},
          doi = {10.3847/1538-4357/acc1e6},
archivePrefix = {arXiv},
       eprint = {2209.06247},
 primaryClass = {astro-ph.GA},
       adsurl = {https://ui.adsabs.harvard.edu/abs/2023ApJ...948..112C}
}

@ARTICLE{Cleri2023ApJ...953...10C,
       author = {{Cleri}, Nikko J. and {Olivier}, Grace M. and {Hutchison}, Taylor A. and {Papovich}, Casey and {Trump}, Jonathan R. and {Amor{\'\i}n}, Ricardo O. and {Backhaus}, Bren E. and {Berg}, Danielle A. and {Fern{\'a}ndez}, Vital and {Finkelstein}, Steven L. and {Fujimoto}, Seiji and {Hirschmann}, Michaela and {Kartaltepe}, Jeyhan S. and {Kocevski}, Dale D. and {Simons}, Raymond C. and {Wilkins}, Stephen M. and {Yung}, L.~Y. Aaron},
        title = "{Using [Ne V]/[Ne III] to Understand the Nature of Extreme-ionization Galaxies}",
      journal = {ApJ},
         year = 2023,
        month = aug,
       volume = {953},
       number = {1},
          eid = {10},
        pages = {10},
          doi = {10.3847/1538-4357/acde55},
archivePrefix = {arXiv},
       eprint = {2301.07745},
 primaryClass = {astro-ph.GA},
       adsurl = {https://ui.adsabs.harvard.edu/abs/2023ApJ...953...10C}
}

@ARTICLE{Cluver2010ApJ...710..248C,
       author = {{Cluver}, M.~E. and {Appleton}, P.~N. and {Boulanger}, F. and {Guillard}, P. and {Ogle}, P. and {Duc}, P. -A. and {Lu}, N. and {Rasmussen}, J. and {Reach}, W.~T. and {Smith}, J.~D. and {Tuffs}, R. and {Xu}, C.~K. and {Yun}, M.~S.},
        title = "{Powerful H$_{2}$ Line Cooling in Stephan's Quintet. I. Mapping the Significant Cooling Pathways in Group-wide Shocks}",
      journal = {ApJ},
         year = 2010,
        month = feb,
       volume = {710},
       number = {1},
        pages = {248-264},
          doi = {10.1088/0004-637X/710/1/248},
archivePrefix = {arXiv},
       eprint = {0912.0282},
 primaryClass = {astro-ph.CO},
       adsurl = {https://ui.adsabs.harvard.edu/abs/2010ApJ...710..248C}
}

@ARTICLE{Cortijo-Ferrero2017A&A...607A..70C,
       author = {{Cortijo-Ferrero}, C. and {Gonz{\'a}lez Delgado}, R.~M. and {P{\'e}rez}, E. and {Cid Fernandes}, R. and {Garc{\'\i}a-Benito}, R. and {Di Matteo}, P. and {S{\'a}nchez}, S.~F. and {de Amorim}, A.~L. and {Lacerda}, E.~A.~D. and {L{\'o}pez Fern{\'a}ndez}, R. and {Tadhunter}, C.},
        title = "{The spatially resolved star formation history of mergers. A comparative study of the LIRGs IC 1623, NGC 6090, NGC 2623, and Mice}",
      journal = {A\&A},
         year = 2017,
        month = nov,
       volume = {607},
          eid = {A70},
        pages = {A70},
          doi = {10.1051/0004-6361/201731217},
archivePrefix = {arXiv},
       eprint = {1707.05324},
 primaryClass = {astro-ph.GA},
       adsurl = {https://ui.adsabs.harvard.edu/abs/2017A&A...607A..70C}
}

@ARTICLE{Croft2006ApJ...647.1040C,
       author = {{Croft}, Steve and {van Breugel}, Wil and {de Vries}, Wim and {Dopita}, Mike and {Martin}, Chris and {Morganti}, Raffaella and {Neff}, Susan and {Oosterloo}, Tom and {Schiminovich}, David and {Stanford}, S.~A. and {van Gorkom}, Jacqueline},
        title = "{Minkowski's Object: A Starburst Triggered by a Radio Jet, Revisited}",
      journal = {\apj},
         year = 2006,
        month = aug,
       volume = {647},
       number = {2},
        pages = {1040-1055},
          doi = {10.1086/505526},
archivePrefix = {arXiv},
       eprint = {astro-ph/0604557},
 primaryClass = {astro-ph},
       adsurl = {https://ui.adsabs.harvard.edu/abs/2006ApJ...647.1040C}
}

@ARTICLE{Dale2014ApJ...784...83D,
       author = {{Dale}, Daniel A. and {Helou}, George and {Magdis}, Georgios E. and {Armus}, Lee and {D{\'\i}az-Santos}, Tanio and {Shi}, Yong},
        title = "{A Two-parameter Model for the Infrared/Submillimeter/Radio Spectral Energy Distributions of Galaxies and Active Galactic Nuclei}",
      journal = {ApJ},
         year = 2014,
        month = mar,
       volume = {784},
       number = {1},
          eid = {83},
        pages = {83},
          doi = {10.1088/0004-637X/784/1/83},
archivePrefix = {arXiv},
       eprint = {1402.1495},
 primaryClass = {astro-ph.GA},
       adsurl = {https://ui.adsabs.harvard.edu/abs/2014ApJ...784...83D}
}

@ARTICLE{DiMatteo2012ApJ...745L..29D,
       author = {{Di Matteo}, T. and {Khandai}, N. and {DeGraf}, C. and {Feng}, Y. and {Croft}, R.~A.~C. and {Lopez}, J. and {Springel}, V.},
        title = "{Cold Flows and the First Quasars}",
      journal = {ApJL},
         year = 2012,
        month = feb,
       volume = {745},
       number = {2},
          eid = {L29},
        pages = {L29},
          doi = {10.1088/2041-8205/745/2/L29},
archivePrefix = {arXiv},
       eprint = {1107.1253},
 primaryClass = {astro-ph.CO},
       adsurl = {https://ui.adsabs.harvard.edu/abs/2012ApJ...745L..29D}
}

@ARTICLE{Draine1993ARA&A..31..373D,
       author = {{Draine}, Bruce T. and {McKee}, Christopher F.},
        title = "{Theory of interstellar shocks.}",
      journal = {Annual Review of Astronomy and Astrophysics},
         year = 1993,
        month = jan,
       volume = {31},
        pages = {373-432},
          doi = {10.1146/annurev.aa.31.090193.002105},
       adsurl = {https://ui.adsabs.harvard.edu/abs/1993ARA&A..31..373D}
}

@ARTICLE{Duan2025MNRAS.540..774D,
       author = {{Duan}, Qiao and {Conselice}, Christopher J. and {Li}, Qiong and {Austin}, Duncan and {Harvey}, Thomas and {Adams}, Nathan J. and {Duncan}, Kenneth J. and {Trussler}, James and {Ferreira}, Leonardo and {Westcott}, Lewi and {Harris}, Honor and {Windhorst}, Rogier A. and {Holwerda}, Benne W. and {Broadhurst}, Thomas J. and {Coe}, Dan and {Cohen}, Seth H. and {Du}, Xiaojing and {Driver}, Simon P. and {Frye}, Brenda and {Grogin}, Norman A. and {Hathi}, Nimish P. and {Jansen}, Rolf A. and {Koekemoer}, Anton M. and {Marshall}, Madeline A. and {Nonino}, Mario and {Ortiz}, III, Rafael and {Pirzkal}, Nor and {Robotham}, Aaron and {Ryan}, Russell E. and {Summers}, Jake and {D'Silva}, Jordan C.~J. and {Willmer}, Christopher N.~A. and {Yan}, Haojing},
        title = "{Galaxy mergers in the epoch of reionization {\textendash} I. A JWST study of pair fractions, merger rates, and stellar mass accretion rates at z = 4.5{\textendash}11.5}",
      journal = {Monthly Notices of the Royal Astronomical Society},
         year = 2025,
        month = jun,
       volume = {540},
       number = {1},
        pages = {774-805},
          doi = {10.1093/mnras/staf638},
archivePrefix = {arXiv},
       eprint = {2407.09472},
 primaryClass = {astro-ph.GA},
       adsurl = {https://ui.adsabs.harvard.edu/abs/2025MNRAS.540..774D}
}

@ARTICLE{Dubois2016MNRAS.463.3948D,
       author = {{Dubois}, Yohan and {Peirani}, S{\'e}bastien and {Pichon}, Christophe and {Devriendt}, Julien and {Gavazzi}, Rapha{\"e}l and {Welker}, Charlotte and {Volonteri}, Marta},
        title = "{The HORIZON-AGN simulation: morphological diversity of galaxies promoted by AGN feedback}",
      journal = {MNRAS},
         year = 2016,
        month = dec,
       volume = {463},
       number = {4},
        pages = {3948-3964},
          doi = {10.1093/mnras/stw2265},
archivePrefix = {arXiv},
       eprint = {1606.03086},
 primaryClass = {astro-ph.GA},
       adsurl = {https://ui.adsabs.harvard.edu/abs/2016MNRAS.463.3948D}
}

@ARTICLE{Duncan2019ApJ...876..110D,
       author = {{Duncan}, Kenneth and {Conselice}, Christopher J. and {Mundy}, Carl and {Bell}, Eric and {Donley}, Jennifer and {Galametz}, Audrey and {Guo}, Yicheng and {Grogin}, Norman A. and {Hathi}, Nimish and {Kartaltepe}, Jeyhan and {Kocevski}, Dale and {Koekemoer}, Anton M. and {P{\'e}rez-Gonz{\'a}lez}, Pablo G. and {Mantha}, Kameswara B. and {Snyder}, Gregory F. and {Stefanon}, Mauro},
        title = "{Observational Constraints on the Merger History of Galaxies since z {\ensuremath{\approx}} 6: Probabilistic Galaxy Pair Counts in the CANDELS Fields}",
      journal = {ApJ},
         year = 2019,
        month = may,
       volume = {876},
       number = {2},
          eid = {110},
        pages = {110},
          doi = {10.3847/1538-4357/ab148a},
archivePrefix = {arXiv},
       eprint = {1903.12188},
 primaryClass = {astro-ph.GA},
       adsurl = {https://ui.adsabs.harvard.edu/abs/2019ApJ...876..110D}
}

@ARTICLE{Ellison2019MNRAS.487.2491E,
       author = {{Ellison}, Sara L. and {Viswanathan}, Akshara and {Patton}, David R. and {Bottrell}, Connor and {McConnachie}, Alan W. and {Gwyn}, Stephen and {Cuillandre}, Jean-Charles},
        title = "{A definitive merger-AGN connection at z {\ensuremath{\sim}} 0 with CFIS: mergers have an excess of AGN and AGN hosts are more frequently disturbed}",
      journal = {MNRAS},
         year = 2019,
        month = aug,
       volume = {487},
       number = {2},
        pages = {2491-2504},
          doi = {10.1093/mnras/stz1431},
archivePrefix = {arXiv},
       eprint = {1905.08830},
 primaryClass = {astro-ph.GA},
       adsurl = {https://ui.adsabs.harvard.edu/abs/2019MNRAS.487.2491E}
}

@ARTICLE{Emonts2014A&A...572A..40E,
       author = {{Emonts}, B.~H.~C. and {Piqueras-L{\'o}pez}, J. and {Colina}, L. and {Arribas}, S. and {Villar-Mart{\'\i}n}, M. and {Pereira-Santaella}, M. and {Garcia-Burillo}, S. and {Alonso-Herrero}, A.},
        title = "{Outflow of hot and cold molecular gas from the obscured secondary nucleus of NGC 3256: closing in on feedback physics}",
      journal = {Astronomy and Astrophysics},
         year = 2014,
        month = dec,
       volume = {572},
          eid = {A40},
        pages = {A40},
          doi = {10.1051/0004-6361/201423805},
archivePrefix = {arXiv},
       eprint = {1409.4468},
 primaryClass = {astro-ph.GA},
       adsurl = {https://ui.adsabs.harvard.edu/abs/2014A&A...572A..40E}
}

@ARTICLE{Emonts2025ApJ...978..111E,
       author = {{Emonts}, B.~H.~C. and {Appleton}, P.~N. and {Lisenfeld}, U. and {Guillard}, P. and {Xu}, C.~K. and {Reach}, W.~T. and {Barcos-Mu{\~n}oz}, L. and {Labiano}, A. and {Ogle}, P.~M. and {O'Sullivan}, E. and {Togi}, A. and {Gallagher}, S.~C. and {Aromal}, P. and {Duc}, P. -A. and {Alatalo}, K. and {Boulanger}, F. and {D{\'\i}az-Santos}, T. and {Helou}, G.},
        title = "{Bird's-eye View of Molecular Gas across Stephan's Quintet Galaxy Group and Intragroup Medium}",
      journal = {\apj},
         year = 2025,
        month = jan,
       volume = {978},
       number = {1},
          eid = {111},
        pages = {111},
          doi = {10.3847/1538-4357/ad957c},
archivePrefix = {arXiv},
       eprint = {2411.14310},
 primaryClass = {astro-ph.GA},
       adsurl = {https://ui.adsabs.harvard.edu/abs/2025ApJ...978..111E}
}

@ARTICLE{Fabian2012ARA&A..50..455F,
       author = {{Fabian}, A.~C.},
        title = "{Observational Evidence of Active Galactic Nuclei Feedback}",
      journal = {ARA\&A},
         year = 2012,
        month = sep,
       volume = {50},
        pages = {455-489},
          doi = {10.1146/annurev-astro-081811-125521},
archivePrefix = {arXiv},
       eprint = {1204.4114},
 primaryClass = {astro-ph.CO},
       adsurl = {https://ui.adsabs.harvard.edu/abs/2012ARA&A..50..455F}
}

@ARTICLE{Fedotov2011AJ....142...42F,
       author = {{Fedotov}, K. and {Gallagher}, S.~C. and {Konstantopoulos}, I.~S. and {Chandar}, R. and {Bastian}, N. and {Charlton}, J.~C. and {Whitmore}, B. and {Trancho}, G.},
        title = "{Star Clusters as Tracers of Interactions in Stephan's Quintet (Hickson Compact Group 92)}",
      journal = {AJ},
         year = 2011,
        month = aug,
       volume = {142},
       number = {2},
          eid = {42},
        pages = {42},
          doi = {10.1088/0004-6256/142/2/42},
archivePrefix = {arXiv},
       eprint = {1105.5840},
 primaryClass = {astro-ph.CO},
       adsurl = {https://ui.adsabs.harvard.edu/abs/2011AJ....142...42F}
}

@ARTICLE{Fogarty2011MNRAS.417..835F,
       author = {{Fogarty}, L. and {Thatte}, N. and {Tecza}, M. and {Clarke}, F. and {Goodsall}, T. and {Houghton}, R. and {Salter}, G. and {Davies}, R.~L. and {Kassin}, S.~A.},
        title = "{SWIFT observations of the Arp 147 ring galaxy system}",
      journal = {Monthly Notices of the Royal Astronomical Society},
         year = 2011,
        month = oct,
       volume = {417},
       number = {2},
        pages = {835-844},
          doi = {10.1111/j.1365-2966.2011.19066.x},
archivePrefix = {arXiv},
       eprint = {1105.4423},
 primaryClass = {astro-ph.GA},
       adsurl = {https://ui.adsabs.harvard.edu/abs/2011MNRAS.417..835F}
}

@ARTICLE{GaiaCollaboration2016A&A...595A...2G,
       author = {{Gaia Collaboration} and {Brown}, A.~G.~A. and {Vallenari}, A. and {Prusti}, T. and {de Bruijne}, J.~H.~J. and {Mignard}, F. and {Drimmel}, R. and {Babusiaux}, C. and {Bailer-Jones}, C.~A.~L. and {Bastian}, U. and {Biermann}, M. and {Evans}, D.~W. and {Eyer}, L. and {Jansen}, F. and {Jordi}, C. and {Katz}, D. and {Klioner}, S.~A. and {Lammers}, U. and {Lindegren}, L. and {Luri}, X. and {O'Mullane}, W. and {Panem}, C. and {Pourbaix}, D. and {Randich}, S. and {Sartoretti}, P. and {Siddiqui}, H.~I. and {Soubiran}, C. and {Valette}, V. and {van Leeuwen}, F. and {Walton}, N.~A. and {Aerts}, C. and {Arenou}, F. and {Cropper}, M. and {H{\o}g}, E. and {Lattanzi}, M.~G. and {Grebel}, E.~K. and {Holland}, A.~D. and {Huc}, C. and {Passot}, X. and {Perryman}, M. and {Bramante}, L. and {Cacciari}, C. and {Casta{\~n}eda}, J. and {Chaoul}, L. and {Cheek}, N. and {De Angeli}, F. and {Fabricius}, C. and {Guerra}, R. and {Hern{\'a}ndez}, J. and {Jean-Antoine-Piccolo}, A. and {Masana}, E. and {Messineo}, R. and {Mowlavi}, N. and {Nienartowicz}, K. and {Ord{\'o}{\~n}ez-Blanco}, D. and {Panuzzo}, P. and {Portell}, J. and {Richards}, P.~J. and {Riello}, M. and {Seabroke}, G.~M. and {Tanga}, P. and {Th{\'e}venin}, F. and {Torra}, J. and {Els}, S.~G. and {Gracia-Abril}, G. and {Comoretto}, G. and {Garcia-Reinaldos}, M. and {Lock}, T. and {Mercier}, E. and {Altmann}, M. and {Andrae}, R. and {Astraatmadja}, T.~L. and {Bellas-Velidis}, I. and {Benson}, K. and {Berthier}, J. and {Blomme}, R. and {Busso}, G. and {Carry}, B. and {Cellino}, A. and {Clementini}, G. and {Cowell}, S. and {Creevey}, O. and {Cuypers}, J. and {Davidson}, M. and {De Ridder}, J. and {de Torres}, A. and {Delchambre}, L. and {Dell'Oro}, A. and {Ducourant}, C. and {Fr{\'e}mat}, Y. and {Garc{\'\i}a-Torres}, M. and {Gosset}, E. and {Halbwachs}, J. -L. and {Hambly}, N.~C. and {Harrison}, D.~L. and {Hauser}, M. and {Hestroffer}, D. and {Hodgkin}, S.~T. and {Huckle}, H.~E. and {Hutton}, A. and {Jasniewicz}, G. and {Jordan}, S. and {Kontizas}, M. and {Korn}, A.~J. and {Lanzafame}, A.~C. and {Manteiga}, M. and {Moitinho}, A. and {Muinonen}, K. and {Osinde}, J. and {Pancino}, E. and {Pauwels}, T. and {Petit}, J. -M. and {Recio-Blanco}, A. and {Robin}, A.~C. and {Sarro}, L.~M. and {Siopis}, C. and {Smith}, M. and {Smith}, K.~W. and {Sozzetti}, A. and {Thuillot}, W. and {van Reeven}, W. and {Viala}, Y. and {Abbas}, U. and {Abreu Aramburu}, A. and {Accart}, S. and {Aguado}, J.~J. and {Allan}, P.~M. and {Allasia}, W. and {Altavilla}, G. and {{\'A}lvarez}, M.~A. and {Alves}, J. and {Anderson}, R.~I. and {Andrei}, A.~H. and {Anglada Varela}, E. and {Antiche}, E. and {Antoja}, T. and {Ant{\'o}n}, S. and {Arcay}, B. and {Bach}, N. and {Baker}, S.~G. and {Balaguer-N{\'u}{\~n}ez}, L. and {Barache}, C. and {Barata}, C. and {Barbier}, A. and {Barblan}, F. and {Barrado y Navascu{\'e}s}, D. and {Barros}, M. and {Barstow}, M.~A. and {Becciani}, U. and {Bellazzini}, M. and {Bello Garc{\'\i}a}, A. and {Belokurov}, V. and {Bendjoya}, P. and {Berihuete}, A. and {Bianchi}, L. and {Bienaym{\'e}}, O. and {Billebaud}, F. and {Blagorodnova}, N. and {Blanco-Cuaresma}, S. and {Boch}, T. and {Bombrun}, A. and {Borrachero}, R. and {Bouquillon}, S. and {Bourda}, G. and {Bouy}, H. and {Bragaglia}, A. and {Breddels}, M.~A. and {Brouillet}, N. and {Br{\"u}semeister}, T. and {Bucciarelli}, B. and {Burgess}, P. and {Burgon}, R. and {Burlacu}, A. and {Busonero}, D. and {Buzzi}, R. and {Caffau}, E. and {Cambras}, J. and {Campbell}, H. and {Cancelliere}, R. and {Cantat-Gaudin}, T. and {Carlucci}, T. and {Carrasco}, J.~M. and {Castellani}, M. and {Charlot}, P. and {Charnas}, J. and {Chiavassa}, A. and {Clotet}, M. and {Cocozza}, G. and {Collins}, R.~S. and {Costigan}, G. and {Crifo}, F. and {Cross}, N.~J.~G. and {Crosta}, M. and {Crowley}, C. and {Dafonte}, C. and {Damerdji}, Y. and {Dapergolas}, A. and {David}, P. and {David}, M. and {De Cat}, P.},
        title = "{Gaia Data Release 1. Summary of the astrometric, photometric, and survey properties}",
      journal = {A\&A},
         year = 2016,
        month = nov,
       volume = {595},
          eid = {A2},
        pages = {A2},
          doi = {10.1051/0004-6361/201629512},
archivePrefix = {arXiv},
       eprint = {1609.04172},
 primaryClass = {astro-ph.IM},
       adsurl = {https://ui.adsabs.harvard.edu/abs/2016A&A...595A...2G}
}

@ARTICLE{Gallagher2001AJ....122..163G,
       author = {{Gallagher}, Sarah C. and {Charlton}, Jane C. and {Hunsberger}, Sally D. and {Zaritsky}, Dennis and {Whitmore}, Bradley C.},
        title = "{Hubble Space Telescope Images of Stephan's Quintet: Star Cluster Formation in a Compact Group Environment}",
      journal = {AJ},
         year = 2001,
        month = jul,
       volume = {122},
       number = {1},
        pages = {163-181},
          doi = {10.1086/321111},
archivePrefix = {arXiv},
       eprint = {astro-ph/0104005},
 primaryClass = {astro-ph},
       adsurl = {https://ui.adsabs.harvard.edu/abs/2001AJ....122..163G}
}

@ARTICLE{Gerber1992ApJ...399L..51G,
       author = {{Gerber}, Richard A. and {Lamb}, Susan A. and {Balsara}, Dinshaw S.},
        title = "{A Model for Ring Galaxies: ARP 147--like Systems}",
      journal = {The Astrophysical Journal},
         year = 1992,
        month = nov,
       volume = {399},
        pages = {L51},
          doi = {10.1086/186604},
       adsurl = {https://ui.adsabs.harvard.edu/abs/1992ApJ...399L..51G}
}

@ARTICLE{Gilli2010A&A...519A..92G,
       author = {{Gilli}, R. and {Vignali}, C. and {Mignoli}, M. and {Iwasawa}, K. and {Comastri}, A. and {Zamorani}, G.},
        title = "{The X-ray to [Ne V]3426 flux ratio: discovering heavily obscured AGN in the distant Universe}",
      journal = {A\&A},
         year = 2010,
        month = sep,
       volume = {519},
          eid = {A92},
        pages = {A92},
          doi = {10.1051/0004-6361/201014039},
archivePrefix = {arXiv},
       eprint = {1005.0373},
 primaryClass = {astro-ph.CO},
       adsurl = {https://ui.adsabs.harvard.edu/abs/2010A&A...519A..92G}
}

@ARTICLE{Grogin2011ApJS..197...35G,
       author = {{Grogin}, Norman A. and {Kocevski}, Dale D. and {Faber}, S.~M. and {Ferguson}, Henry C. and {Koekemoer}, Anton M. and {Riess}, Adam G. and {Acquaviva}, Viviana and {Alexander}, David M. and {Almaini}, Omar and {Ashby}, Matthew L.~N. and {Barden}, Marco and {Bell}, Eric F. and {Bournaud}, Fr{\'e}d{\'e}ric and {Brown}, Thomas M. and {Caputi}, Karina I. and {Casertano}, Stefano and {Cassata}, Paolo and {Castellano}, Marco and {Challis}, Peter and {Chary}, Ranga-Ram and {Cheung}, Edmond and {Cirasuolo}, Michele and {Conselice}, Christopher J. and {Roshan Cooray}, Asantha and {Croton}, Darren J. and {Daddi}, Emanuele and {Dahlen}, Tomas and {Dav{\'e}}, Romeel and {de Mello}, Du{\'\i}lia F. and {Dekel}, Avishai and {Dickinson}, Mark and {Dolch}, Timothy and {Donley}, Jennifer L. and {Dunlop}, James S. and {Dutton}, Aaron A. and {Elbaz}, David and {Fazio}, Giovanni G. and {Filippenko}, Alexei V. and {Finkelstein}, Steven L. and {Fontana}, Adriano and {Gardner}, Jonathan P. and {Garnavich}, Peter M. and {Gawiser}, Eric and {Giavalisco}, Mauro and {Grazian}, Andrea and {Guo}, Yicheng and {Hathi}, Nimish P. and {H{\"a}ussler}, Boris and {Hopkins}, Philip F. and {Huang}, Jia-Sheng and {Huang}, Kuang-Han and {Jha}, Saurabh W. and {Kartaltepe}, Jeyhan S. and {Kirshner}, Robert P. and {Koo}, David C. and {Lai}, Kamson and {Lee}, Kyoung-Soo and {Li}, Weidong and {Lotz}, Jennifer M. and {Lucas}, Ray A. and {Madau}, Piero and {McCarthy}, Patrick J. and {McGrath}, Elizabeth J. and {McIntosh}, Daniel H. and {McLure}, Ross J. and {Mobasher}, Bahram and {Moustakas}, Leonidas A. and {Mozena}, Mark and {Nandra}, Kirpal and {Newman}, Jeffrey A. and {Niemi}, Sami-Matias and {Noeske}, Kai G. and {Papovich}, Casey J. and {Pentericci}, Laura and {Pope}, Alexandra and {Primack}, Joel R. and {Rajan}, Abhijith and {Ravindranath}, Swara and {Reddy}, Naveen A. and {Renzini}, Alvio and {Rix}, Hans-Walter and {Robaina}, Aday R. and {Rodney}, Steven A. and {Rosario}, David J. and {Rosati}, Piero and {Salimbeni}, Sara and {Scarlata}, Claudia and {Siana}, Brian and {Simard}, Luc and {Smidt}, Joseph and {Somerville}, Rachel S. and {Spinrad}, Hyron and {Straughn}, Amber N. and {Strolger}, Louis-Gregory and {Telford}, Olivia and {Teplitz}, Harry I. and {Trump}, Jonathan R. and {van der Wel}, Arjen and {Villforth}, Carolin and {Wechsler}, Risa H. and {Weiner}, Benjamin J. and {Wiklind}, Tommy and {Wild}, Vivienne and {Wilson}, Grant and {Wuyts}, Stijn and {Yan}, Hao-Jing and {Yun}, Min S.},
        title = "{CANDELS: The Cosmic Assembly Near-infrared Deep Extragalactic Legacy Survey}",
      journal = {ApJS},
         year = 2011,
        month = dec,
       volume = {197},
       number = {2},
          eid = {35},
        pages = {35},
          doi = {10.1088/0067-0049/197/2/35},
archivePrefix = {arXiv},
       eprint = {1105.3753},
 primaryClass = {astro-ph.CO},
       adsurl = {https://ui.adsabs.harvard.edu/abs/2011ApJS..197...35G}
}

@ARTICLE{Guillard2009A&A...502..515G,
       author = {{Guillard}, P. and {Boulanger}, F. and {Pineau Des For{\^e}ts}, G. and {Appleton}, P.~N.},
        title = "{H$_{2}$ formation and excitation in the Stephan's Quintet galaxy-wide collision}",
      journal = {A\&A},
         year = 2009,
        month = aug,
       volume = {502},
       number = {2},
        pages = {515-528},
          doi = {10.1051/0004-6361/200811263},
archivePrefix = {arXiv},
       eprint = {0904.4239},
 primaryClass = {astro-ph.CO},
       adsurl = {https://ui.adsabs.harvard.edu/abs/2009A&A...502..515G}
}

@ARTICLE{Guillard2012ApJ...749..158G,
       author = {{Guillard}, P. and {Boulanger}, F. and {Pineau des For{\^e}ts}, G. and {Falgarone}, E. and {Gusdorf}, A. and {Cluver}, M.~E. and {Appleton}, P.~N. and {Lisenfeld}, U. and {Duc}, P. -A. and {Ogle}, P.~M. and {Xu}, C.~K.},
        title = "{Turbulent Molecular Gas and Star Formation in the Shocked Intergalactic Medium of Stephan's Quintet}",
      journal = {ApJ},
         year = 2012,
        month = apr,
       volume = {749},
       number = {2},
          eid = {158},
        pages = {158},
          doi = {10.1088/0004-637X/749/2/158},
archivePrefix = {arXiv},
       eprint = {1202.2862},
 primaryClass = {astro-ph.CO},
       adsurl = {https://ui.adsabs.harvard.edu/abs/2012ApJ...749..158G}
}

@ARTICLE{Hani2018MNRAS.475.1160H,
       author = {{Hani}, Maan H. and {Sparre}, Martin and {Ellison}, Sara L. and {Torrey}, Paul and {Vogelsberger}, Mark},
        title = "{Galaxy mergers moulding the circum-galactic medium - I. The impact of a major merger}",
      journal = {MNRAS},
         year = 2018,
        month = mar,
       volume = {475},
       number = {1},
        pages = {1160-1176},
          doi = {10.1093/mnras/stx3252},
archivePrefix = {arXiv},
       eprint = {1801.06183},
 primaryClass = {astro-ph.GA},
       adsurl = {https://ui.adsabs.harvard.edu/abs/2018MNRAS.475.1160H}
}

@ARTICLE{Hardcastle2020NewAR..8801539H,
       author = {{Hardcastle}, M.~J. and {Croston}, J.~H.},
        title = "{Radio galaxies and feedback from AGN jets}",
      journal = {NewAR},
         year = 2020,
        month = jun,
       volume = {88},
          eid = {101539},
        pages = {101539},
          doi = {10.1016/j.newar.2020.101539},
archivePrefix = {arXiv},
       eprint = {2003.06137},
 primaryClass = {astro-ph.HE},
       adsurl = {https://ui.adsabs.harvard.edu/abs/2020NewAR..8801539H}
}

@ARTICLE{Hasinger2018ApJ...858...77H,
       author = {{Hasinger}, G. and {Capak}, P. and {Salvato}, M. and {Barger}, A.~J. and {Cowie}, L.~L. and {Faisst}, A. and {Hemmati}, S. and {Kakazu}, Y. and {Kartaltepe}, J. and {Masters}, D. and {Mobasher}, B. and {Nayyeri}, H. and {Sanders}, D. and {Scoville}, N.~Z. and {Suh}, H. and {Steinhardt}, C. and {Yang}, Fengwei},
        title = "{The DEIMOS 10K Spectroscopic Survey Catalog of the COSMOS Field}",
      journal = {ApJ},
         year = 2018,
        month = may,
       volume = {858},
       number = {2},
          eid = {77},
        pages = {77},
          doi = {10.3847/1538-4357/aabacf},
archivePrefix = {arXiv},
       eprint = {1803.09251},
 primaryClass = {astro-ph.GA},
       adsurl = {https://ui.adsabs.harvard.edu/abs/2018ApJ...858...77H}
}

@ARTICLE{Hickson1982ApJ...255..382H,
       author = {{Hickson}, P.},
        title = "{Systematic properties of compact groups of galaxies.}",
      journal = {ApJ},
         year = 1982,
        month = apr,
       volume = {255},
        pages = {382-391},
          doi = {10.1086/159838},
       adsurl = {https://ui.adsabs.harvard.edu/abs/1982ApJ...255..382H}
}

@ARTICLE{Hopkins2006ApJS..163....1H,
       author = {{Hopkins}, Philip F. and {Hernquist}, Lars and {Cox}, Thomas J. and {Di Matteo}, Tiziana and {Robertson}, Brant and {Springel}, Volker},
        title = "{A Unified, Merger-driven Model of the Origin of Starbursts, Quasars, the Cosmic X-Ray Background, Supermassive Black Holes, and Galaxy Spheroids}",
      journal = {ApJS},
         year = 2006,
        month = mar,
       volume = {163},
       number = {1},
        pages = {1-49},
          doi = {10.1086/499298},
archivePrefix = {arXiv},
       eprint = {astro-ph/0506398},
 primaryClass = {astro-ph},
       adsurl = {https://ui.adsabs.harvard.edu/abs/2006ApJS..163....1H}
}

@ARTICLE{Hunter2023PASP..135g4501H,
       author = {{Hunter}, Todd R. and {Indebetouw}, Remy and {Brogan}, Crystal L. and {Berry}, Kristin and {Chang}, Chin-Shin and {Francke}, Harold and {Geers}, Vincent C. and {G{\'o}mez}, Laura and {Hibbard}, John E. and {Humphreys}, Elizabeth M. and {Kent}, Brian R. and {Kepley}, Amanda A. and {Kunneriath}, Devaky and {Lipnicky}, Andrew and {Loomis}, Ryan A. and {Mason}, Brian S. and {Masters}, Joseph S. and {Maud}, Luke T. and {Muders}, Dirk and {Sabater}, Jose and {Sugimoto}, Kanako and {Sz{\H{u}}cs}, L{\'a}szl{\'o} and {Vasiliev}, Eugene and {Videla}, Liza and {Villard}, Eric and {Williams}, Stewart J. and {Xue}, Rui and {Yoon}, Ilsang},
        title = "{The ALMA Interferometric Pipeline Heuristics}",
      journal = {\pasp},
         year = 2023,
        month = jul,
       volume = {135},
       number = {1049},
          eid = {074501},
        pages = {074501},
          doi = {10.1088/1538-3873/ace216},
archivePrefix = {arXiv},
       eprint = {2306.07420},
 primaryClass = {astro-ph.IM},
       adsurl = {https://ui.adsabs.harvard.edu/abs/2023PASP..135g4501H}
}

@ARTICLE{Izotov2012MNRAS.427.1229I,
       author = {{Izotov}, Y.~I. and {Thuan}, T.~X. and {Privon}, G.},
        title = "{The detection of [Ne V] emission in five blue compact dwarf galaxies}",
      journal = {MNRAS},
         year = 2012,
        month = dec,
       volume = {427},
       number = {2},
        pages = {1229-1237},
          doi = {10.1111/j.1365-2966.2012.22051.x},
archivePrefix = {arXiv},
       eprint = {1209.5265},
 primaryClass = {astro-ph.CO},
       adsurl = {https://ui.adsabs.harvard.edu/abs/2012MNRAS.427.1229I}
}

@ARTICLE{Jin2018ApJ...864...56J,
       author = {{Jin}, Shuowen and {Daddi}, Emanuele and {Liu}, Daizhong and {Smol{\v{c}}i{\'c}}, Vernesa and {Schinnerer}, Eva and {Calabr{\`o}}, Antonello and {Gu}, Qiusheng and {Delhaize}, Jacinta and {Delvecchio}, Ivan and {Gao}, Yu and {Salvato}, Mara and {Puglisi}, Annagrazia and {Dickinson}, Mark and {Bertoldi}, Frank and {Sargent}, Mark and {Novak}, Mladen and {Magdis}, Georgios and {Aretxaga}, Itziar and {Wilson}, Grant W. and {Capak}, Peter},
        title = "{{\textquotedblleft}Super-deblended{\textquotedblright} Dust Emission in Galaxies. II. Far-IR to (Sub)millimeter Photometry and High-redshift Galaxy Candidates in the Full COSMOS Field}",
      journal = {ApJ},
         year = 2018,
        month = sep,
       volume = {864},
       number = {1},
          eid = {56},
        pages = {56},
          doi = {10.3847/1538-4357/aad4af},
archivePrefix = {arXiv},
       eprint = {1807.04697},
 primaryClass = {astro-ph.GA},
       adsurl = {https://ui.adsabs.harvard.edu/abs/2018ApJ...864...56J}
}

@ARTICLE{Kennicutt2012ARA&A..50..531K,
       author = {{Kennicutt}, Robert C. and {Evans}, Neal J.},
        title = "{Star Formation in the Milky Way and Nearby Galaxies}",
      journal = {ARA\&A},
         year = 2012,
        month = sep,
       volume = {50},
        pages = {531-608},
          doi = {10.1146/annurev-astro-081811-125610},
archivePrefix = {arXiv},
       eprint = {1204.3552},
 primaryClass = {astro-ph.GA},
       adsurl = {https://ui.adsabs.harvard.edu/abs/2012ARA&A..50..531K}
}

@ARTICLE{Kim2010ApJ...724..386K,
       author = {{Kim}, Dohyeong and {Im}, Myungshin and {Kim}, Minjin},
        title = "{New Estimators of Black Hole Mass in Active Galactic Nuclei with Hydrogen Paschen Lines}",
      journal = {The Astrophysical Journal},
         year = 2010,
        month = nov,
       volume = {724},
       number = {1},
        pages = {386-399},
          doi = {10.1088/0004-637X/724/1/386},
archivePrefix = {arXiv},
       eprint = {1012.1112},
 primaryClass = {astro-ph.CO},
       adsurl = {https://ui.adsabs.harvard.edu/abs/2010ApJ...724..386K}
}

@ARTICLE{Koekemoer2011ApJS..197...36K,
       author = {{Koekemoer}, Anton M. and {Faber}, S.~M. and {Ferguson}, Henry C. and {Grogin}, Norman A. and {Kocevski}, Dale D. and {Koo}, David C. and {Lai}, Kamson and {Lotz}, Jennifer M. and {Lucas}, Ray A. and {McGrath}, Elizabeth J. and {Ogaz}, Sara and {Rajan}, Abhijith and {Riess}, Adam G. and {Rodney}, Steve A. and {Strolger}, Louis and {Casertano}, Stefano and {Castellano}, Marco and {Dahlen}, Tomas and {Dickinson}, Mark and {Dolch}, Timothy and {Fontana}, Adriano and {Giavalisco}, Mauro and {Grazian}, Andrea and {Guo}, Yicheng and {Hathi}, Nimish P. and {Huang}, Kuang-Han and {van der Wel}, Arjen and {Yan}, Hao-Jing and {Acquaviva}, Viviana and {Alexander}, David M. and {Almaini}, Omar and {Ashby}, Matthew L.~N. and {Barden}, Marco and {Bell}, Eric F. and {Bournaud}, Fr{\'e}d{\'e}ric and {Brown}, Thomas M. and {Caputi}, Karina I. and {Cassata}, Paolo and {Challis}, Peter J. and {Chary}, Ranga-Ram and {Cheung}, Edmond and {Cirasuolo}, Michele and {Conselice}, Christopher J. and {Roshan Cooray}, Asantha and {Croton}, Darren J. and {Daddi}, Emanuele and {Dav{\'e}}, Romeel and {de Mello}, Duilia F. and {de Ravel}, Loic and {Dekel}, Avishai and {Donley}, Jennifer L. and {Dunlop}, James S. and {Dutton}, Aaron A. and {Elbaz}, David and {Fazio}, Giovanni G. and {Filippenko}, Alexei V. and {Finkelstein}, Steven L. and {Frazer}, Chris and {Gardner}, Jonathan P. and {Garnavich}, Peter M. and {Gawiser}, Eric and {Gruetzbauch}, Ruth and {Hartley}, Will G. and {H{\"a}ussler}, Boris and {Herrington}, Jessica and {Hopkins}, Philip F. and {Huang}, Jia-Sheng and {Jha}, Saurabh W. and {Johnson}, Andrew and {Kartaltepe}, Jeyhan S. and {Khostovan}, Ali A. and {Kirshner}, Robert P. and {Lani}, Caterina and {Lee}, Kyoung-Soo and {Li}, Weidong and {Madau}, Piero and {McCarthy}, Patrick J. and {McIntosh}, Daniel H. and {McLure}, Ross J. and {McPartland}, Conor and {Mobasher}, Bahram and {Moreira}, Heidi and {Mortlock}, Alice and {Moustakas}, Leonidas A. and {Mozena}, Mark and {Nandra}, Kirpal and {Newman}, Jeffrey A. and {Nielsen}, Jennifer L. and {Niemi}, Sami and {Noeske}, Kai G. and {Papovich}, Casey J. and {Pentericci}, Laura and {Pope}, Alexandra and {Primack}, Joel R. and {Ravindranath}, Swara and {Reddy}, Naveen A. and {Renzini}, Alvio and {Rix}, Hans-Walter and {Robaina}, Aday R. and {Rosario}, David J. and {Rosati}, Piero and {Salimbeni}, Sara and {Scarlata}, Claudia and {Siana}, Brian and {Simard}, Luc and {Smidt}, Joseph and {Snyder}, Diana and {Somerville}, Rachel S. and {Spinrad}, Hyron and {Straughn}, Amber N. and {Telford}, Olivia and {Teplitz}, Harry I. and {Trump}, Jonathan R. and {Vargas}, Carlos and {Villforth}, Carolin and {Wagner}, Cory R. and {Wandro}, Pat and {Wechsler}, Risa H. and {Weiner}, Benjamin J. and {Wiklind}, Tommy and {Wild}, Vivienne and {Wilson}, Grant and {Wuyts}, Stijn and {Yun}, Min S.},
        title = "{CANDELS: The Cosmic Assembly Near-infrared Deep Extragalactic Legacy Survey{\textemdash}The Hubble Space Telescope Observations, Imaging Data Products, and Mosaics}",
      journal = {ApJS},
         year = 2011,
        month = dec,
       volume = {197},
       number = {2},
          eid = {36},
        pages = {36},
          doi = {10.1088/0067-0049/197/2/36},
archivePrefix = {arXiv},
       eprint = {1105.3754},
 primaryClass = {astro-ph.CO},
       adsurl = {https://ui.adsabs.harvard.edu/abs/2011ApJS..197...36K}
}

@ARTICLE{Kristensen2023A&A...675A..86K,
       author = {{Kristensen}, L.~E. and {Godard}, B. and {Guillard}, P. and {Gusdorf}, A. and {Pineau des For{\^e}ts}, G.},
        title = "{Shock excitation of H$_{2}$ in the James Webb Space Telescope era}",
      journal = {Astronomy and Astrophysics},
         year = 2023,
        month = jul,
       volume = {675},
          eid = {A86},
        pages = {A86},
          doi = {10.1051/0004-6361/202346254},
archivePrefix = {arXiv},
       eprint = {2307.04178},
 primaryClass = {astro-ph.GA},
       adsurl = {https://ui.adsabs.harvard.edu/abs/2023A&A...675A..86K}
}

@ARTICLE{Kron1980ApJS...43..305K,
       author = {{Kron}, R.~G.},
        title = "{Photometry of a complete sample of faint galaxies.}",
      journal = {ApJS},
         year = 1980,
        month = jun,
       volume = {43},
        pages = {305-325},
          doi = {10.1086/190669},
       adsurl = {https://ui.adsabs.harvard.edu/abs/1980ApJS...43..305K}
}

@ARTICLE{Li2024ApJS..275...27L,
       author = {{Li}, Mingyu and {Zhang}, Haibin and {Cai}, Zheng and {Liang}, Yongming and {Kashikawa}, Nobunari and {Ma}, Ke and {Fan}, Xiaohui and {Prochaska}, J. Xavier and {Emonts}, Bjorn H.~C. and {Wang}, Xin and {Wu}, Yunjing and {Zhang}, Shiwu and {Li}, Qiong and {Johnson}, Sean D. and {Yue}, Minghao and {Arrigoni Battaia}, Fabrizio and {Cantalupo}, Sebastiano and {Hennawi}, Joseph F. and {Kikuta}, Satoshi and {Ning}, Yuanhang and {Ouchi}, Masami and {Shimakawa}, Rhythm and {Wang}, Ben and {Wang}, Weichen and {Zheng}, Zheng and {Zheng}, Zhen-Ya},
        title = "{MAMMOTH-Subaru. II. Diverse Populations of Circumgalactic Ly{\ensuremath{\alpha}} Nebulae at Cosmic Noon}",
      journal = {ApJS},
         year = 2024,
        month = dec,
       volume = {275},
       number = {2},
          eid = {27},
        pages = {27},
          doi = {10.3847/1538-4365/ad812c},
archivePrefix = {arXiv},
       eprint = {2405.13113},
 primaryClass = {astro-ph.GA},
       adsurl = {https://ui.adsabs.harvard.edu/abs/2024ApJS..275...27L}
}

@ARTICLE{Li2024arXiv240514980L,
       author = {{Li}, Junyao and {Zhuang}, Ming-Yang and {Shen}, Yue and {Volonteri}, Marta and {Chen}, Nianyi and {Di Matteo}, Tiziana},
        title = "{Active Galactic Nuclei and Host Galaxies in COSMOS-Web. II. First Look at the Kpc-scale Dual and Offset AGN Population}",
      journal = {arXiv e-prints},
         year = 2024,
        month = may,
          eid = {arXiv:2405.14980},
        pages = {arXiv:2405.14980},
          doi = {10.48550/arXiv.2405.14980},
archivePrefix = {arXiv},
       eprint = {2405.14980},
 primaryClass = {astro-ph.GA},
       adsurl = {https://ui.adsabs.harvard.edu/abs/2024arXiv240514980L}
}

@BOOK{Osterbrock2006agna.book.....O,
       author = {{Osterbrock}, Donald E. and {Ferland}, Gary J.},
        title = "{Astrophysics of gaseous nebulae and active galactic nuclei}",
         year = 2006,
       adsurl = {https://ui.adsabs.harvard.edu/abs/2006agna.book.....O}
}
\bibliographystyle{aasjournalv7}



\end{document}